\documentclass[
reprint,
superscriptaddress,
showkeys,
amsmath,
amssymb,
aps,
floatfix,
]{revtex4-2}

\usepackage{graphicx}
\usepackage{dcolumn}
\usepackage{bm}
\usepackage{float}
\usepackage{tabularray}
\usepackage{longtable}

\usepackage{natbib}[sort&compress]
\usepackage{xcolor}

\usepackage{color}
\usepackage{soul}
\usepackage{amssymb, nicefrac}
\usepackage[most]{tcolorbox}
\usepackage{hyperref}
\hypersetup{
	colorlinks=true,
	linkcolor=blue,
	filecolor=magenta,      
	urlcolor=blue,
	citecolor=blue,
	pdfpagemode=FullScreen,
}

\usepackage{cleveref}
\crefname{figure}{Fig.}{Figs.}
\crefname{equation}{Eq.}{Eqs.}

\begin{document}
	\title{EVALUATION OF FISSION FRAGMENT MOMENTS OF INERTIA FOR SPONTANEOUS FISSION OF CF-252}
	\author{D.E. Lyubashevsky}
	\email[]{lyubashevskiy@phys.vsu.ru}
	\affiliation{Voronezh State University, Voronezh, Russia}
	\affiliation{International Institute of Computer Technologies, Voronezh, Russia}
	\author{P.V. Kostryukov}
	\affiliation{Voronezh State University of Forestry and Technologies, Voronezh, Russia}
	\affiliation{Voronezh State University, Voronezh, Russia}
	\author{A.A. Pisklyukov}
	\affiliation{Voronezh State University, Voronezh, Russia}
	\author{J.D. Shcherbina}
	\affiliation{Voronezh State University, Voronezh, Russia}
	
	\date{\today}
	
	\begin{abstract}
	 Current work discusses methods for estimating the moments of inertia of fission fragments for spontaneous fission of the isotope Cf-252, in particular, two main approaches are mentioned: statistical and microscopic. In addition, the methods of the classical and superfluid approaches to the calculation of the moments of inertia are discussed, as well as their application to different models of nuclei. Within this framework, the influence of different oscillation modes and nucleon exchange on the moments of inertia and spin distributions of fission fragments is evaluated. The authors emphasizes the need for a comparative analysis of theoretical predictions with experimental data for a deeper understanding of the internal structure of nuclei and fission mechanisms.
	\end{abstract}
	
	
	\maketitle
	
	\section{Introduction}\label{sec: intro}
	
	Understanding how the spins of fission products are formed is still a puzzle in the nuclear fission theory. This is particularly interesting when the spin of the parent nucleus is zero or very small, while the spins of the fission fragments (FFs) can reach values of up to six or seven $\hbar$ units. Currently, there are a couple of candidate models~\cite{stetcu2021, randrup&vogt2021} that can explain this effect. At the same time, they are not without limitations and drawbacks, which argues for additional research aimed at developing a general and consistent model to describe the given phenomenon. 

	The first model proposed by Randrup's group is based on a statistical model \texttt{FREYA}~\cite{randrup2009,verbeke2018} using the Monte Carlo method to determine the mass, charge, and velocity of primary fission fragments (PFFs). It follows the ideas of~\cite{dossing1985_I, dossing1985_II}, where the importance of the influence of nucleon exchange on the dynamical evolution of PFF spins was shown, leading to the excitation of six rotational modes (transverse wriggling and bending modes, which are doubly degenerate, and longitudinal twisting and tilting modes), and the mobility coefficients, which determine the time scales for these oscillations, were estimated. Taking advantage of the results of this study in \texttt{FREYA} the realization of a complex mechanism of nucleon exchange, assumes complete relaxation of transverse modes (wriggling and bending), while longitudinal modes (twisting and tilting) are not excited~\cite{vogt2013,randrup2014}.
	
	The second model proposed by Bulgac's group~\cite{stetcu2021,bulgac2021,bulgac2022} is based on a microscopic approach using time-dependent density functional theory (TDDFT) and constructed by analogy with the derivation of Fermi's golden rule. Since this model is based on rather general considerations, it can be concluded from the analysis mentioned in the papers that the basic approach is quantum angular momentum theory, together with some assumptions on the individual spin and angular momentum distributions. In the framework of this approach, contributions to the FF spins not only from transverse bending and wriggling oscillations, but also from longitudinal tilting and twisting oscillations have been taken into account, allowing to describe the spin in the three-dimensional case, which should affect, for example, the angular spin distributions and other important characteristics. The three-dimensional model of the spins in the article~\cite{bulgac2022} differs significantly from the two-dimensional phenomenological model~\cite{vogt2021}, where the spin is described in planes perpendicular to the $Z$-axis.
	
	These two different model representations also have some similarities. For example, in \cite{bulgac2022} the importance of considering the FFs moments of inertia is discussed, as well as in \cite{randrup&vogt2021}, where it is noted that this quantity plays a key role in the description of the FFs spin distributions, since it reflects their nucleon structure, and introduced a special parameterization for them. The authors called it ``at hot'', by which they obtained a sawtooth behavior of the given moments as a function of the fragment mass number, and hypothesized that the behavior will be the same for the FF spins during fission.
	
	Nevertheless, despite significant progress in the development of these models, each of them faces certain difficulties and limitations. The model proposed by Bulgak's group, although based on the fundamental principles of quantum mechanics, might lead to unpredictable or partially incorrect results, especially in the absence of empirical data. On the other hand, phenomenological models, such as the one implemented in the \texttt{FREYA}, require a large number of parameters and knowledge of nuclear properties that are typically not available or precise enough. A good example here is the introduced ``at hot'' approximation has no physical justification, although it leads to a reasonable agreement between experimental and calculated by \texttt{FREYA} spins.
	
	These challenges emphasize the importance of analyzing and comparing theoretical predictions with experimental data. Therefore, the aim of the present work is to perform a comparative analysis of the FFs moments of inertia calculated using the most advanced and accurate theoretical approaches at heavy actinide nuclei case. Particular attention is paid not only to evaluating their agreement with experimental data, but also to studying the possibility of predicting the spin distribution of the FF. In addition, an important aspect of the analysis will be to test the hypothesis of the sawtooth character of the momentum of inertia variation as a function of the FF mass number, which will provide a deeper understanding of the internal structure of the nuclei and their fission mechanisms.
	
	\section{ESTIMATION METHODS MOMENT OF INERTIA}\label{sec: inertia calculation approaches}
	
	Let us assume that an axially symmetric nucleus slowly rotates with a certain frequency $\omega$ around an axis perpendicular to the axis of symmetry of the nucleus $Z$. In this case, the nucleus can be represented as an ellipsoid of rotation, the semi-axis lengths of which are equal to
	\begin{equation} \label{eq:1}
		R_1=\left(1-\frac{1}{2}\sqrt{\frac{5}{4\pi}}\beta \right)R, \quad R_2 =\left(1+\sqrt{\frac{5}{4 \pi}}\beta \right)R,
	\end{equation}
	where $R$ is the axis length and $\beta$ is the deformation parameter. 
	The angular momentum is then related to the rotation of the nucleus by the relation: 
	\begin{equation} \label{eq:2}
		\hbar R= \omega J,
	\end{equation}
	at here $\hbar$ is the reduced Planck constant, $J$ the moment of inertia. 
	
	The latter is particularly important because it reflects the internal structure of the nucleus. There are several methods for its determination, based on different concepts, which we will consider below.  
	
	\subsection{Classical approaches}\label{sbs:hydrodynamical approach}
	
	The rotation of the nucleus can be considered through the classical representations of the rotation of a rigid body or the potential motion of an ideal fluid in a rotating shell, which are limiting cases of the rotational model of the nucleus. Their difference is essential, which can be shown by a simple example of the rotation of a spherical body. In the first approach, the moment of inertia tends to a finite value - the moment of inertia of a spherically symmetric body. In the other, the velocity of each point of the surface is tangential, and thus the normal component of the velocity will be zero. Thus it follows~\cite{sitenko2014} from the equations of hydrodynamics for an ideal fluid to satisfy a resting fluid, and so the moment of inertia of the system is zero. 

\begin{figure}[h]
		\centering
		\includegraphics[width=0.8\linewidth]{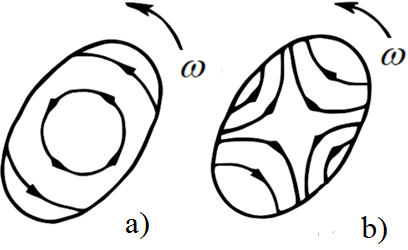}
		\vspace{-1em}
		\caption{The velocity field in the case of the rotation of a solid body (a) and the potential motion of an ideal fluid in a rotating rigid shell (b).}\label{fig:1}
	\end{figure}

	For the non-spherical case, the normal component of the velocity on the surface is already different from zero, and the fluid will be entrained in the rotation of the shell. It means that the rotational energy for a given angular velocity will be greater the more the shape of the shell differs from the sphere. In such a case, let us determine the moment of inertia for the potential motion of the fluid in the core. The motion is described by a potential $\varphi$ that obeys the Laplace equation: $\Delta\varphi = 0$, and the velocity of the fluid $v$ is determined by the potential gradient. For an ideal fluid, the boundary conditions reduce to the requirement that the normal component of the fluid velocity at the surface coincides with the normal component of the velocity at the vessel wall. In other words, while in the case of a solid body the entire system rotates as a whole, in the case of a rigid shell filled with an ideal fluid, the fluid is only entrained by the walls near the surface of the shell. A good illustration of the behavior of these velocity fields is~\cref{fig:1}

	When the nucleus is considered as a solid body, the moment of inertia has the form
	\begin{equation} \label{eq:3}
		J_0 \!=\! \frac{m}{5}\left(R_1^2 \!+\! R_2^2\right)\! =\! \frac{2}{5}mR^2  \left( 1 \!+\!  \frac{1}{2}\sqrt{\frac{5}{4\pi}}\beta \!+\! \frac{25}{32\pi}\beta^2\right),
	\end{equation}
	where $m$ is the mass of the nucleus.
	While in the hydrodynamic model of the nucleus~\cite{sitenko2014} the moment of inertia of the nucleus is represented as
	\begin{equation}\label{eq:4}
	J=\frac{m}{5}\frac{\left(R_2^2-R_1^2\right)^2}{R_2^2+R_1^2}
	\end{equation}
	which using the expression~\eqref{eq:2}, can be rewritten as
	\begin{equation}\label{eq:5}
		J=\frac{9mR^2}{4\pi} \frac{\beta^2 \left(1+\frac{1}{4}\sqrt{\frac{5}{4\pi}}\beta\right)^2}{2+\sqrt{\frac{5}{4\pi}}\beta +\frac{25}{16\pi}\beta^2}.
	\end{equation}
	When the deformation parameter $\beta$ is small, both models have a relationship which is determined as follows:
	\begin{equation}\label{eq:6}
		\frac{J}{J_0}=\frac{45 \beta^2\left(1 + \frac{\beta}{4}\sqrt{\frac{5}{4\pi}}\right)}{8 \pi \! \left(2 \!+\! \sqrt{\frac{5}{4\pi}}\beta \! +\! \frac{25}{16 \pi}\beta^2 \right) \! \left(1 \! + \! \sqrt{\frac{5}{16\pi}}\beta \! + \! \frac{25}{32\pi}\beta^2 \right)}.
	\end{equation}
	Such a representation of the moment of inertia is very convenient to use in the framework of the model of independent particles moving in a non-spherical well, for example in an anisotropically oscillating one, where its behavior is similar to the moment of inertia of a rigid body~\eqref{eq:3}. The reason for this is that during the slow adiabatic rotation of the potential well, the states of the separate, non-correlated particles do not change, so the system rotates as a solid.

	Nevertheless, the observed moments of inertia of the nuclei differ from the solid-state values, as shown by the rich experimental base. The deviations are caused by residual interactions between nucleons, which slow down the collective rotation and reduce the moment of inertia of the system. If the interaction between the particles were significant, their mean path length would be smaller than the size of the nucleus, which is closer to the behavior of the hydrodynamic model, in which the collective motion of the nucleons becomes potential. If the interaction between nucleons is due to pairing strengths, then a superfluid model of the nucleus must be applied. The consistent method for calculating the moments of inertia of nuclei on the basis of the mentioned model was developed by Migdal~\cite{migdal1960}, the main points of which are presented below. 

	\subsection{Superfluid nucleus approach}\label{sbs:superfluid approach}

	The atomic nucleus is a purely quantum system, so the above solid-state and hydrodynamic approaches to estimating of moments of inertia require corrections. The superfluid model of the nucleus, as well as the many-particle shell model, assumes that the nucleons in the nucleus move as predicted by the single-particle shell model, and takes into account the residual interaction, i.e. the correlations between the nucleons. In the superfluid model, however, the residual interaction is taken into account by quite different, more sophisticated methods. The residual interaction can be distinguished from a specific short-range pairing interaction, which plays a crucial role in explaining the properties of nuclei. Some experimental data (e.g. the binding energy of the last neutron in light nuclei, the zero spin of even-even nuclei in the ground state, etc.) indicate that a strong correlation between two nucleons is realized only when these nucleons are in states of identical energy and quantum number, except for the projections of their total moments. The individual nucleons in the nucleus are assumed to have the same single-particle states as in the independent-particle model, so the pairing of nucleons can be described by the quantum numbers of the aforementioned model.

	The superfluidity of nuclei is related to the fact that nucleons form pairs due to the residual interaction between them. These pairs are created by a special type of correlations called superconducting pair correlations. As a result, the ground state energy of the nucleus is much lower than it would be without these correlations, and such a state of the nucleus is called superfluid.

	The effects of correlated pairs and superfluidity affect the moments of inertia of the nuclei. For example, due to the superfluid properties of nuclei, their values can be 2-3 times smaller than in analogous calculations for the solid state model. Therefore, the moments of inertia of nuclei will be calculated in the framework of the superfluidity theory developed for homogeneous unbounded Fermi systems in works~\cite{bogoliubov1958,tolmachev1958}. 

	Using Migdal’s approach~\cite{migdal1960}, we calculate the moment of inertia for the oscillatory potential:
	\begin{equation} \label{eq:7}
		J=J_0\left\{1-g_1+\frac{g_1^2\chi^2}{v_1^2 g_1 + g^2_2\nu_2} \right\} = J_0 \Phi_1\left( \chi  \right).
	\end{equation}
	The values of the function $\Phi_1(\chi)$ are given in~\cref{tab:1} (for $\nu_2 = 10$), and the parameter $\chi$ is defined by the expression
	\begin{equation}\label{eq:8}
		\chi =\frac{\varepsilon_0}{\Delta p_0 R_0} \cdot \beta,
	\end{equation}
	where $\beta =\frac{2\left( a - b \right)}{a + b}$ is the deformation parameter of the nucleus, $a$ and $b$ are the semi-axes of the spheroid; $R_0=\frac{a+b}{2}$; $p_0$ is the momentum operator of the particle; $\varepsilon_0$ is the energy of the particle; $\Delta$ is the mass defect.

	Now in the framework of~\cite{migdal1960} let us calculate the moment of inertia for a rectangular potential well, introducing the variables $\eta = \frac{m}{l}{l}$, $\xi = \frac{l}{l_0}$, wherefrom one obtains
	\begin{equation}\label{eq:9}
		\begin{aligned}
			J_1 & \!=\!J_0 \! \left\{\! 1 \!- \! \frac{45}{4} \! \int\limits_0^1 \! d\xi \xi^3\sqrt{1 \! - \! \xi ^2} \! \int\limits_0^1 \! d\eta \left(1 \! - \! \eta^2 \right) \! g \! \left(\! \aleph \frac{\eta}{\xi}  \! \right) \! \right\} \\
			&= J_0 \Phi_2 (\chi).
		\end{aligned}
	\end{equation}
	Values of the function $\Phi_2 (\chi)$ are also given in~\cref{tab:1}.
		
	\begin{table*}
		\caption{\label{tab:1}The values of the functions $\Phi _1\left( \chi \right)$ and $\Phi_2\left( \chi \right)$ defined within the superfluid model of work~\cite{migdal1960}}
		\begin{ruledtabular}
			\centering
			\begin{tabular}{cccccccccc}
				$\chi$ & $\Phi_1 \left( \chi \right)$ & $\Phi_2 \left(\chi \right)$ & $\chi$ & $\Phi_1 \left( \chi \right)$ & $\Phi_2 \left(\chi \right)$ & $\chi$ & $\Phi_1 \left( \chi \right)$ & $\Phi_2 \left(\chi \right)$\\
				\hline
				0 & 0  & 0 & 0.9 & 0.43  & 0.17 & 1.8 & 0.77 & 0.34\\
				0.1 & 0.01  & 0.01 & 1.0 & 0.49  & 0.19 & 1.9 & 0.79 & 0.36\\
				0.2 & 0.03  & 0.02 & 1.1 & 0.53  & 0.21 & 2.0 & 0.80 & 0.37\\
				0.3 & 0.07  & 0.03 & 1.2 & 0.60  & 0.23 & 2.2 & 0.81 & 0.40\\
				0.4 & 0.13  & 0.06 & 1.3 & 0.64  & 0.24 & 2.4 & 0.83 & 0.43\\
				0.5 & 0.19  & 0.08 & 1.4 & 0.67  & 0.26 & 2.6 & 0.85 & 0.45\\
				0.6 & 0.26  & 0.10 & 1.5 & 0.71  & 0.28 & 2.8 & 0.86 & 0.48\\
				0.7 & 0.32  & 0.13 & 1.6 & 0.74  & 0.30 & 3.0 & 0.87 & 0.50\\
				0.8 & 0.38  & 0.15 & 1.7 & 0.75 & 0.32 &  &  &\\ 
			\end{tabular}
		\end{ruledtabular}
	\end{table*}
	
	The calculations in~\cref{tab:1} were performed for only one type of nucleon, but for comparison with experimental data both types must be considered. When $Z > 20$, the Fermi levels of neutrons and protons diverge, making neutron-proton pairing impossible. From this point on, the nuclei consist of two fluids, neutron and proton, which cannot exchange the momentum of motion due to the presence of an energy gap in the excitations of each fluid. In this case, the moment of inertia of the nucleus is represented as the sum of the moments of inertia of the neutrons and protons. Denoting the ratio $J_k / {J_0}_k = \Phi(\chi_k)$, we obtain the expression
	\begin{equation} \label{eq:10}
		\frac{J}{J_0}=\frac{N}{A}\Phi(\chi_n) + \frac{Z}{A}\Phi(\chi_p),
	\end{equation}
	at here $\chi_n$ and $\chi_p$ are values for neutrons and protons, respectively. 

	The $\Delta$ values included in the definition~\eqref{eq:8} cannot be calculated theoretically and must be taken from experiment. However, from the definition of the Green's function we can obtain
	\begin{equation}\label{eq:11}
		\begin{aligned}
			G_{\lambda}=\sum\limits_s \frac{\left| \Phi _{N+1}^s, a_{\lambda}^{+}\Phi_N^0 \right|^2}{\varepsilon - E_s (N+1) + E_0 (N) + i\delta} & \\
			+\sum\limits_s \frac{\left| \Phi _{N-1}^s, a_{\lambda}\Phi _N^0 \right|^2}{\varepsilon +E_s (N-1) - E_0(N) - i\delta}&,
		\end{aligned}
	\end{equation}
	where $a_{\lambda}$, $a_{\lambda}^+$ are the particle annihilation and birth operators in the $\lambda$ state. 

\begin{figure}[h]
		\centering
		\includegraphics[width=\linewidth]{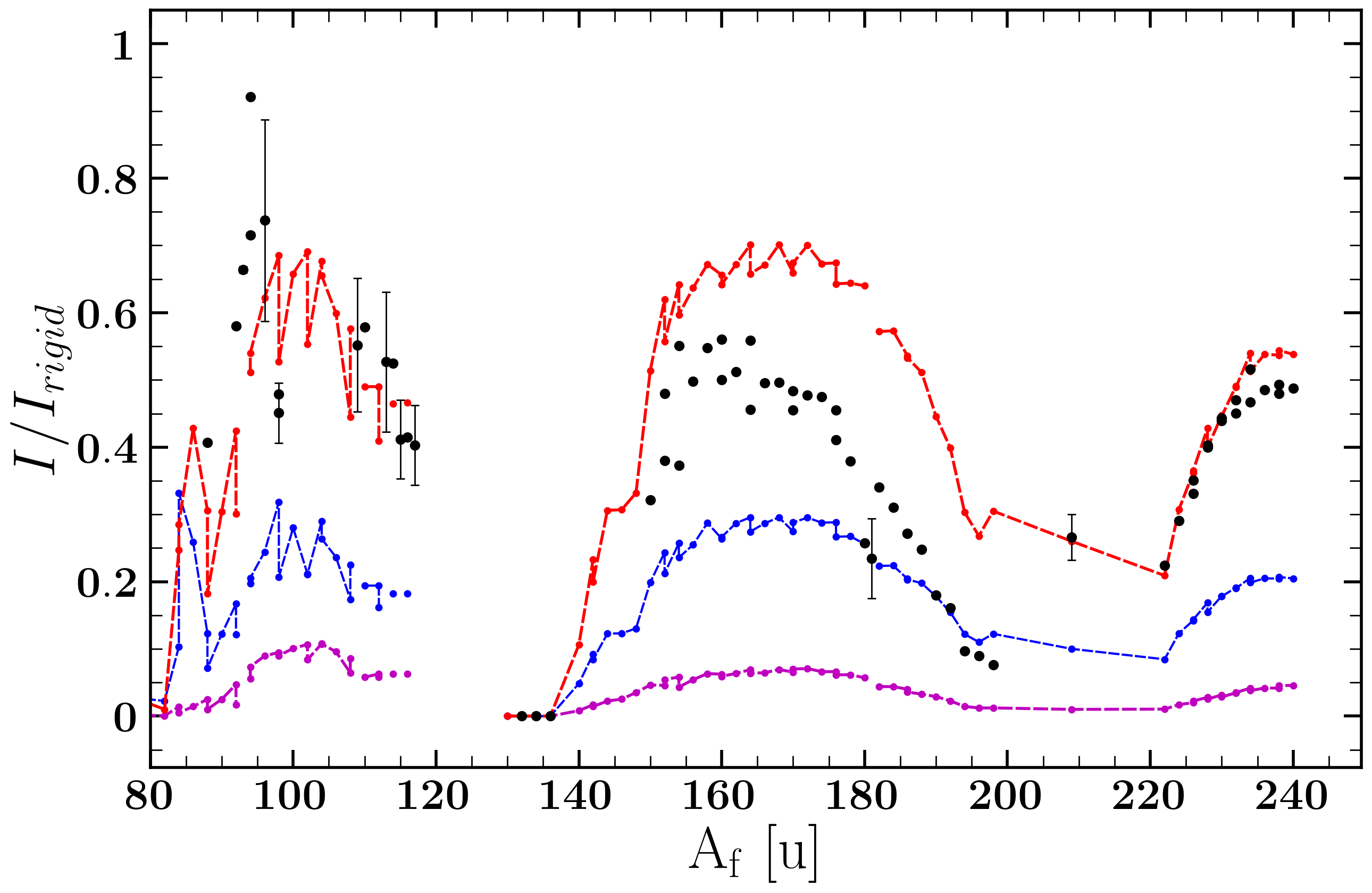}
		\vspace{-1.5em}
		\caption{Comparison of the moments of inertia of nuclei as a function of their mass number $A_f$ calculated within the hydrodynamic approach (magenta) and the superfluid core model for the cases of rectangular (blue) and oscillatory Black dots denote experimental data taken from~\cite{lovchikova1983, sheikh2020}. All values are normalized to the values of the solid state moments.}\label{fig:2}
	\end{figure}

	Comparing the formula \eqref{eq:11} with the expression from \cite{migdal1960}, it follows that
	\begin{equation} \label{eq:12}
		\begin{aligned}
			& E_0 (N+1) + E_0(N) = \Delta + \varepsilon_0, \\ 
			& E_0 (N) - E_0 (N - 1)= -\Delta + \varepsilon_0.
		\end{aligned}	
	\end{equation}
	These formulas allow us to find $\Delta$ (MeV) from the binding energy of nuclei. A more accurate expression can be obtained by excluding from $E_0(N)$ the non-pairing dependence on $N$. For this purpose, let us write out an expression in which the first and second derivative terms in the $N' - N$ expansion are excluded. This condition is satisfied by the value
	\begin{equation}\label{eq:13}
		\frac{1}{4}[3E (N+1) - 3E (N) + E (N-1) - E (N+2)] = \Delta.
	\end{equation}
	To compare~\cref{eq:7,eq:8} with the experimental moments of inertia, we express $\aleph$ and $\chi$ through the observed values. We obtain $R_0 = R \left(1 + \frac{1}{3} \beta \right)$. Given that $R = 1.2 A^{1/3} \cdot 10^{-13}$ cm, we find that
	\begin{equation} \label{eq:14}
		\begin{aligned}
			&\varepsilon_{0}^n=52 \frac{M}{M_{eff}} \left( \frac{N}{A} \right)^{2/3}\, , \\
			&p_{0}^{n}R= 1.9 \cdot {N^{1/3}}\, , \\
			&\chi_n=\frac{\beta}{1+\frac{\beta}{3}}\frac{27}{\Delta_n A^{1/3}}\left(\frac{N}{A} \right)^{1/3}.
	\end{aligned}
	\end{equation}
	Similar expressions have the form for protons:
	\begin{equation} \label{eq:15}
	\begin{aligned}
		&\varepsilon_0^p = 52 \frac{M}{M_{eff}} \left(\frac{Z}{A} \right)^{2/3} \;, \\
		&p_0^p R=1.9\cdot Z^{1/3} \;, \\
		&\chi_p=\frac{\beta}{1+\frac{\beta}{3}}\frac{27}{\Delta_p A^{1/3}}\left(\frac{Z}{A} \right)^{1/3}.
	\end{aligned}
	\end{equation}
	Using~\cref{eq:3,eq:5,eq:14,eq:15}, the values of the moments of inertia of the nuclei were obtained. They are normalized to the solid moment of inertia and are presented in~\cref{fig:2} and Table~\ref{tab:2}.
	From the presented figure it is easy to see that the best agreement between the experimental data and the different theoretical approaches is achieved within the framework of the oscillatory potential superfluid nucleus model. This is to be expected, since this approach takes into account the interaction between nucleons, which slows down the collective rotation, leading to a decrease in the moment of inertia of the system with respect to its solid-state counterpart. The observed overestimation of the moments of inertia of equilibrium deformed fragments in the 150 to 200 mass range is probably due to the use of oscillatory and rectangular well potentials, which is a serious simplification. The use of more realistic or complex types of potentials would improve the description. 
	
	We have considered several methods for estimating moments of inertia that provide more accurate theoretical predictions. These approaches provide insight into the dynamics of nuclear fission and the nature of the pre-fragment moments of inertia. However, they did not take into account that they are highly deformed during fission, and there is currently no direct experimental method capable of accurately estimating their quadrupole nonequilibrium deformation parameters. 
	
	Nevertheless, the authors of this paper have developed an indirect method to estimate these deformations. Before proceeding with this method, however, it is necessary to consider some important assumptions and concepts concerning low-energy fission that will be used in the further analysis.
	
	\section{Behavior of pre-fragments' moments of inertia at the scission point}\label{sec:moments behavior}
	
	\subsection{Non-equilibrium deformation determination of pre-fragments}\label{sbs:non-equilibrium deformations}
	
	In understanding the main characteristics of the mentioned type of fission, the main role is played by quantum interference effects described in the framework of quantum fission theory, in particular, the authors refer to the works~\cite{bohr1939, nix1965, bohr_mottelson, sushkov1982,tanimura1987,barabanov1997,kadmensky2002,kadmensky2004,kadmensky2005}. In it, nucleonic, collective deformation and oscillation modes of motion of the fissile nucleus associated with the evolution of its shape up to the scission into PFF are taken into account simultaneously. 
	
	Thus, the characteristics of spontaneous nuclear fission can be described within the framework of two fundamental assumptions proposed in the work~\cite{bohr_mottelson}. First, at all stages of the fission process, the compound nucleus preserves its axial symmetry, which does not contradict the existing experimental data. Second, at the stage of descent from the outer saddle point to the scission point, the projection $K$ of the spin $J$ onto its symmetry axis $Z$ is conserved, i.e. $K$ remains an integral of motion. The latter has its own difficulties, since the transient fissile state with the corresponding $K$ can be sensitive to heating, both of the fissile nucleus before its scission and of the PFFs themselves in early stages of their evolution. This is due to the fact that at significant heating ($\approx 1 $ MeV) of an axial-symmetric nucleus, the effect of dynamical amplification of the Coriolis interaction occurs~\cite{kadmensky2002}, leading to a uniform statistical mixing of all possible values of $K$ projections at rather small values of spin $J$. In this case, all $K$ projections near the scission point become equally probable, leading~\cite{bunakov2016,kadmensky2017} to the complete disappearance of asymmetries, including different $P$ and $T$ parities, in the angular distributions of the products of binary and ternary fission. However, such anisotropies are observed experimentally~\cite{carruthers1968,jesinger2002,danilyan2009,danilyan2010,valsky2010,gagarskii2011,guseva2013,danilyan2019} even in the case of induced low-energy fission, where the excitation energies are higher than in spontaneous fission. It follows that within the framework of the considered type of fission, the composite nucleus remains "cold" at all stages, including the process of descent from the outer saddle point and the formation of PFF angular distributions. 

	For this reason, within the framework of the ``coldness'' hypothesis of the fissile nucleus, we assume that all the excitation energy of the fission pre-fragments is ``pumped'' into collective deformation states, leading to their non-equilibrium deformation. After the breakup of the compound fissile system, the already heated FFs are de-excited at times $\tau_{nuc}$ by neutron emission, from which the excitation energy of the FFs can be determined. For example, using the results of Strutinsky~\cite{strutinsky1967}, where the relationship between the excitation energy and the deformation of the fragments is established in the framework of the Liquid Drop Model (LDM), we can estimate the value we are looking for. In other words, the non-equilibrium pre-fragment deformations can be related to the average number of emitted neutrons from the specified FFs. 
	
	The question then arises as to the method of performing the neutron yield estimation. In this work, two approaches are used.
	
	The first approach is the numerical calculation of the average number of emitted neutrons in the program package \texttt{FREYA}~\cite{randrup2009,hagmann2016,verbeke2018}. After the previously mentioned stages of PFF generation and nucleon exchange, the FF thermalization process is reproduced by successive neutron emission, which occurs until the excitation energy of the FF is below the threshold $S_n + Q_{min}$, where $S_n$ is the binding energy of the neutron in the FF and $Q_{min} = 0.01$ MeV, i.e. the emission process occurs as long as it is energetically possible. When the neutron emission capability is exhausted, the excited FFs release the remaining excitation by emitting gamma quanta.
	
	The second approach, developed in the paper~\cite{grudzevich2000}, is based on the theoretical evaluation of the decay of the FF taking into account the conservation laws of total angular momentum and parity.  As the author correctly notes, it is very time-consuming, since it is necessary to set a large amount of initial data, primarily data on the characteristics of the excitation: level density and discrete level diagrams. The situation is aggravated by the fact that FFs are neutron-excess nuclei, and it is a non-trivial task to obtain experimental information on the neutron resonance density. The solution was found by using a data library developed under the auspices of the IAEA by an international group of experts for theoretical calculations of nuclear reaction cross sections~\cite{iaea_handbook1998}. However, it included only those nuclei for which neutron resonance densities are known, and the data for discrete levels performed on the basis of NUDAT (version of February 23, 1996) contained a number of both fundamental and technical errors. Therefore, the Grudzewicz group created a new library LDPL-98 for density parameters and discrete level diagrams, which included information on more than 2000 nuclei far from the stability line and allowed to correctly estimate spectra and cross sections of these nuclei. Another difference in the approach was that neutron yields were calculated only for those FFs where the ratio of the experimental value of a given fragment to the maximum neutron yield was at least $0.1$. In~\cite{hagmann2016}, all nuclei for which neutron multiplicity data were available were considered.

\begin{figure}[h]
		\centering
		\includegraphics[width=\linewidth]{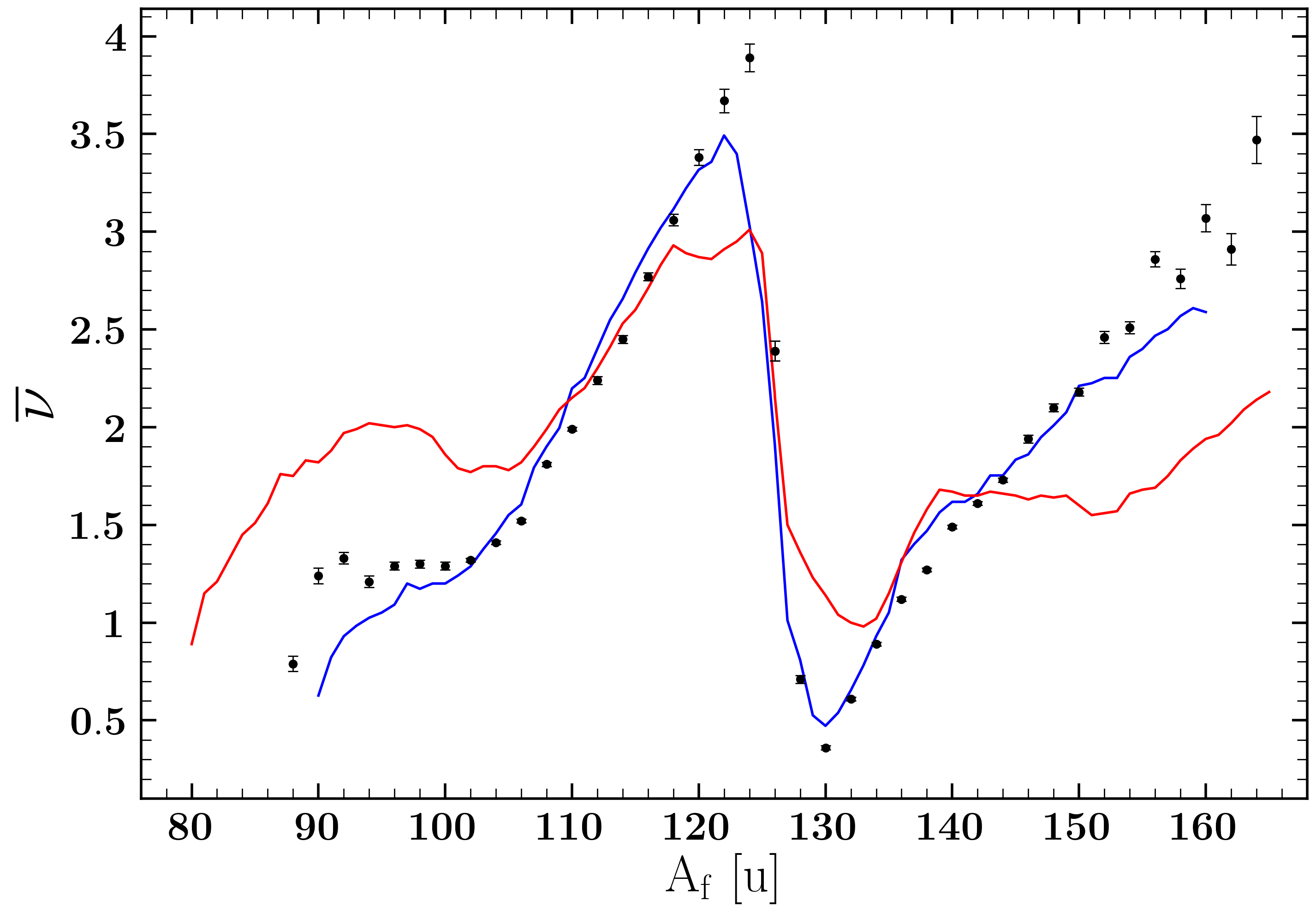}
		\vspace{-1.5em}
		\caption{Mean neutron multiplicities as a function of the mass of spontaneous fission fragments $\rm^{252}Cf$. The black dots indicate experimental data from the work~\cite{walsh1977}, the red line is the \texttt{FREYA} model estimate, and the blue line is the Grudzevich~\cite{grudzevich2000} method evaluation.}\label{fig:3}
	\end{figure}

	On~\cref{fig:3} are presented results of calculations using the approaches described above and available experimental data from the~\cite{walsh1977} study. Within the framework of the theoretical models, multiplicities have been calculated for FFs with mass numbers between 80 and 160. It is not difficult to see that the calculations performed in \texttt{FREYA} are qualitatively similar to the experiment, but the approach of~\cite{grudzevich2000} is in better agreement with the experimental values. 

	Next we determine the excitation energy $U$. For this purpose using the formula proposed in \cite{grudzevich2000} for the relation between the excitation energy $U$ and the neutron multiplicity, as follows
	\begin{equation} \label{eq:16}
		U=5+4\nu +{\nu }^2,
	\end{equation}
	where $\nu$ is the neutron multiplicity.	
	The nonlinearity of the function may indicate that the average energy taken away by neutrons increases with their number as the remaining nucleus approaches the stability line.
	
\begin{figure}[h]
			\centering
			\includegraphics[width=\linewidth]{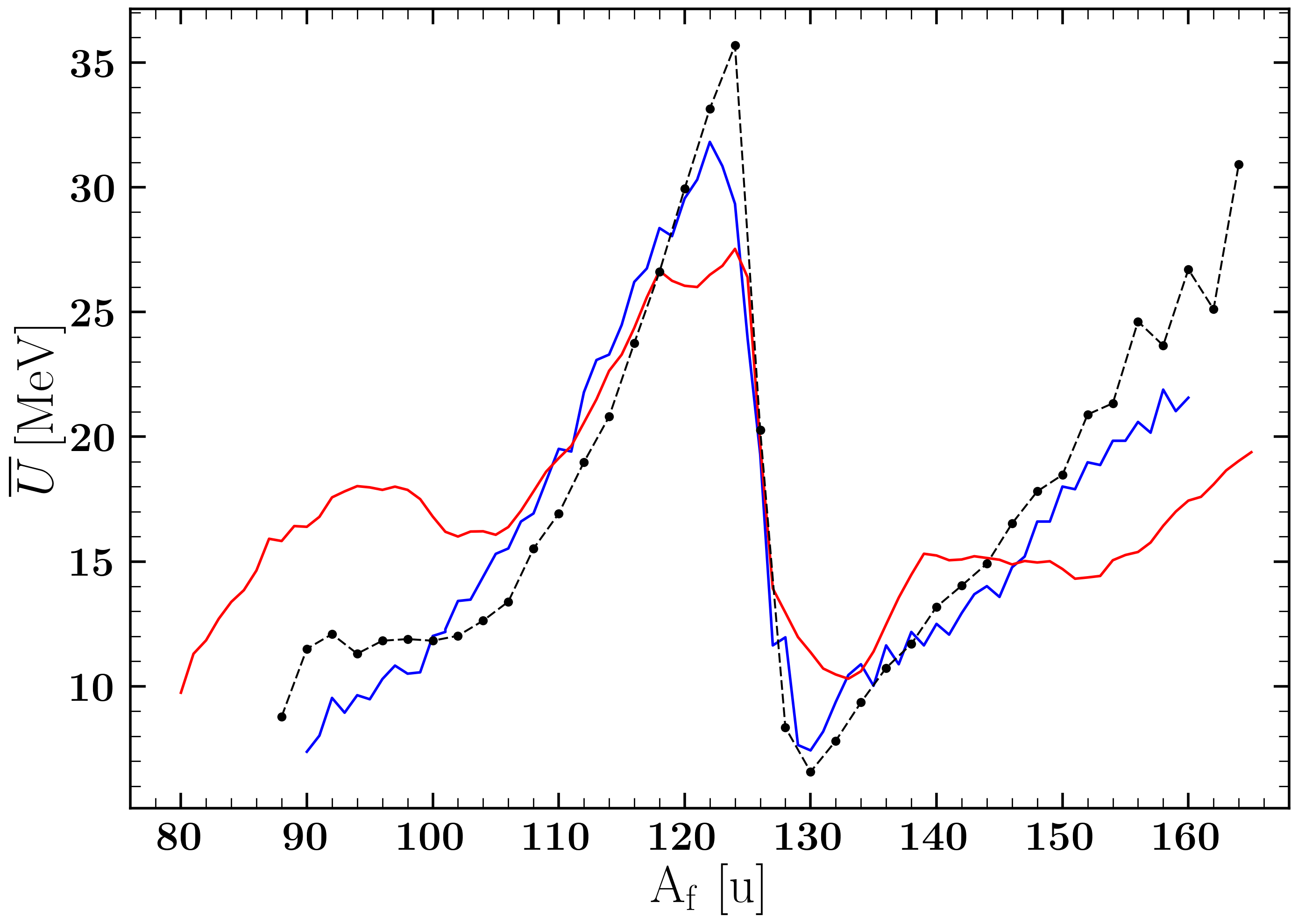}
			\vspace{-1.5em}
			\caption{Mean excitation energies as a function of spontaneous fission fragment mass $\rm^{252}Cf$. Black dots with dashed line denote the reconstructed excitation energies from the data of work~\cite{walsh1977}, red line denotes \texttt{FREYA}, blue line denotes the estimation by the method of Grudzevich's work~\cite{grudzevich2000}}\label{fig:4}
		\end{figure}

	Applying~\cref{eq:16}, we construct the averaged excitation energies $\overline{U}$ shown in~\cref{fig:4}, the red line is the line obtained from the FREYA, while the blue line corresponds to the data calculated in~\cite{grudzevich2000}, and the black dots indicate the energies derived from the experimental data~\cite{walsh1977}.  From the comparative analysis~\cref{fig:4} one can see a reasonable agreement of the considered approaches -- the sawtooth structure is preserved, and in all approaches for both neutron yield multiplicity and excitation energy.

	It only remains to establish the relationship between the excitation energy of the FF and their non-equilibrium deformation. For this purpose, as already mentioned, we will use the method proposed in~\cite{strutinsky1967}. Strutinsky suggested using the Nilsson level scheme to calculate the shell correction to the energy obtained in the LDM as a function of occupation number and deformation. He found a strong correlation between the shell correction and the density of nucleon levels at the Fermi energy. In this case, suppose that the excitation energy $U$ is represented as the deformation energy calculated in the LDM framework, which is the sum of the surface and Coulomb energies, which can be written in simple form as
	\begin{equation}\label{eq:17}
	U=\sigma A^{2/3}\left[ 0.4 ( 1 - x )\alpha^2 - 0.0381( 1 - 2 x ) \alpha^3 \right],
	\end{equation}
	where coefficients are $\sigma =16$ MeV, $x= Z^2 / \left( 45 \cdot A \right)$, and the value $\alpha$ is related to the deformation parameter by the ratio $\alpha =\frac{2}{3}\beta$.
	
	Using~\cref{eq:17} we can calculate the equilibrium deformation energy by substituting the equilibrium deformation values taken from~\cite{moller2016} into the above formula. Since neither the \texttt{FREYA}, nor the results of Grudzevich's work are in full agreement with the experimental data~\cite{walsh1977}, it seems most reasonable to obtain the excitation energies from them using the expression~\eqref{eq:16}. By adding the excitation energy calculated from the evaluation to the equilibrium values of the deformation energy, we obtain the non-equilibrium excitation energy. Then, given the data and using~\cref{eq:17}, the inverse problem is solved, from which the non-equilibrium deformations of the fission pre-fragments are determined. All these values are given in \cref{tab:2}. 
	
	\begin{table*}
	\caption{Values ${\Delta }_n$ and ${\Delta }_p$, calculated from mass defects for some elements\label{tab:2}}
	\begin{ruledtabular}
		\centering
		\begin{tabular}{ccccccccc}
		Nuclei & ${\Delta}_n$ & ${\Delta}_p$ & ${\chi}_n$ &  ${\chi}_p$ & $\beta $ & $J^{osc}_{sup}$ & $J^{rec}_{sup}$ & $J_{hyd}$ \\
		\hline
		${\rm ^{94}Sr}$  & $-$ 0.98 & $-$1.43  & $-$2.54 & $-$1.53 & 0.588 & 7.76 & 6.33 & 5.35 \\
		${\rm ^{158}Nd}$  & $-$0.64 & $-$0.91 & $-$4.64 & $-$2.76 & 0.892 & 12.10 & 10.50 & 9.10 \\ \hline
		${\rm ^{96}Sr}$ & $-$1.03 & $-$1.40 & $-$2.58 & $-$1.65 &	0.639 &  7.78 & 6.25 & 5.29\\
		${\rm ^{156}Nd}$ & $-$0.67 & $-$0.89 & $-$4.35	& $-$2.80 & 0.867 & 11.82 & 10.08 & 8.66\\ \hline
		${\rm ^{98}Sr}$	& $-$0.73 & $-$1.297 & $-$3.66 & $-$1.77 & 0.645 & 7.97 & 6.81 & 5.25 \\
		${\rm ^{154}Nd}$ & $-$0.69 & $-$0.62 & $-$4.21 & $-$4.03 & 0.864 & 11.87 & 10.88 & 8.53\\ \hline
		${\rm ^{100}Zr}$ & $-$0.90 & $-$1.34 & $-$3.12 & $-$1.83 & 0.697 & 8.14 & 6.10 & 5.52 \\
		${\rm ^{152}Sm}$ & $-$1.12 & $-$1.12 & $-$2.49 & $-$2.12 & 0.813 & 11.28 & 8.34 & 8.28\\ \hline
		${\rm ^{102}Zr}$ & $-$0.97 & $-$1.24 & $-$3.02 & $-$2.02 & 0.740 & 8.27 & 6.38 & 5.49\\
		${\rm ^{150}Ce}$ & $-$0.79 & $-$1.03 & $-$3.35& $-$2.20 & 0.759 & 11.25 & 8.84 & 7.59\\ \hline
		${\rm ^{104}Zr}	$ & $-$0.93 & $-$1.27 & $-$3.16	& $-$1.97 &	0.740 & 8.27 & 6.30 & 5.49 \\
		${\rm ^{148}Ce}$ & $-$1.00 & $-$1.27 & $-$2.65 & $-$1.80 & 0.759 & 10.77 & 8.07 & 7.39 \\ \hline
		${\rm ^{106}Mo}$ & $-$1.04 & $-$1.45 & $-$2.84 & $-$1.78 & 0.763 & 8.19	& 6.23 & 5.50 \\
		${\rm ^{146}Ba}$ & $-$0.95 & $-$1.23 & $-$2.68 & $-$1.67 & 0.715 & 10.54 & 7.89 & 6.76\\ \hline
		${\rm ^{108}Mo}$ & $-$1.06 & $-$1.43 & $-$2.69 & $-$1.73 & 0.775 & 8.28 & 6.21 &5.40 \\
		${\rm ^{144}Ba}$ & $-$0.92 & $-$1.19 & $-$2.63 & $-$1.74 & 0.668 & 10.10 & 7.36 & 7.47\\ \hline
		${\rm ^{110}Ru}$ & $-$1.19 & $-$1.34 & $-$2.58 & $-$2.00 & 0.810 & 8.81 & 6.50 & 5.99\\
		${\rm ^{142}Xe}$ & $-$0.94 & $-$1.13 & $-$2.42 & $-$1.41 & 0.615 & 10.20 & 7.41 & 5.70\\ \hline
		${\rm ^{112}Ru}$ & $-$1.18 & $-$1.26 & $-$2.68& $-$2.18	& 0.845	&9.81 & 7.36 & 6.86\\
		${\rm ^{140}Xe}$ & $-$0.98 & $-$1.22 & $-$2.30 & $-$1.58 & 0.604 & 10.79 & 7.51 & 5.93\\ \hline
		${\rm ^{114}Pd}$ & $-$1.36 & $-$1.37 & $-$2.29 & $-$2.10 & 0.900 & 9.58 & 7.04 & 7.36\\
		${\rm ^{138}Te}$ & $-$0.88 & $-$1.14 & $-$3.09 & $-$2.01 & 0.567 & 11.10 & 8.59 & 7.02\\ \hline
		${\rm ^{116}Pd}$ & $-$1.30 & $-$1.35 & $-$2.47 & $-$2.18 & 0.938 & 9.74	& 7.06 & 7.70\\
		${\rm ^{136}Te}$ & $-$0.83 & $-$1.12 & $-$2.57& $-$1.62	& 0.559 & 10.56 & 7.70 & 5.04\\ \hline
		${\rm ^{130}Sn}$ & $-$1.14 & $-$1.70 & $-$1.56 & $-$0.89 & 0.444 & 8.27 & 5.29 & 3.75\\
		${\rm ^{122}Cd}$ & $-$1.30 & $-$1.34 & $-$2.67 & $-$2.25 & 0.997 & 10.58 & 8.01 & 8.07\\
		\hline
		${\rm ^{132}Sn}$ & $-$1.55 & $-$1.85 & $-$1.26 & $-$0.89 & 0.498 & 8.00 & 4.94 &4.29\\
		${\rm ^{120}Cd}$ &$-$1.32 & $-$1.32 & $-$2.60 & $-$2.27 & 0.976 & 10.32 & 7.69 & 8.08\\ \hline
		${\rm ^{134}Te}$ & $-$1.48 & $-$0.94 & $-$1.42 & $-$1.93 & 0.549 & 9.45 & 6.19 & 4.91\\
		${\rm ^{118}Pd}$ & $-$1.23 & $-$1.34 & $-$2.69 & $-$2.13 & 0.914 & 10.07 & 7.51 & 7.48\\
		\hline
		${\rm ^{136}Te}$ & $-$0.83 & $-$1.12 & $-$2.57 & $-$1.62 & 0.559 & 10.79 & 7.82 & 5.31 \\
		${\rm ^{116}Pd}$ & $-$1.30 & $-$1.35 & $-$2.60 & $-$2.18 & 0.938 & 10.56 & 7.88 & 7.88 \\ 
		\hline
		${\rm ^{138}Xe}$ & $-$0.90 & $-$1.18 & $-$2.46 & $-$1.62 & 0.591 & 10.90 & 7.77 & 5.82 \\
		${\rm ^{114}Ru}$ & $-$1.17 & $-$1.20 & $-$2.75 & $-$2.30 & 0.865 & 10.24 & 7.72 & 7.23 \\
		\hline
		${\rm ^{140}Xe}$ & $-$0.98 & $-$1.22 & $-$2.29 & $-$1.58 & 0.604 & 11.08 & 7.89 & 6.10 \\
		${\rm ^{112}Ru}$ & $-$1.18 & $-$1.27 & $-$2.68 & $-$2.15 & 0.845 & 10.07 & 7.51 & 7.00 \\
		\hline
		${\rm ^{142}Xe}$ & $-$0.94 & $-$1.13 & $-$2.42 & $-$1.71 & 0.615 & 10.36 & 7.54 & 5.82\\
		${\rm ^{110}Ru}$ & $-$1.19 & $-$1.34 & $-$2.63 & $-$2.01 & 0.817 & 9.09 & 6.71 & 6.15\\
		\hline
		${\rm ^{144}Ba}$ & $-$0.92 & $-$1.19 & $-$2.63 & $-$1.74 & 0.668 & 10.24 & 7.55 & 6.21\\
		${\rm ^{108}Mo}$ & $-$1.06 & $-$1.44 & $-$2.83 & $-$1.79 & 0.775 & 8.40 & 6.26 & 5.58\\ 
		\hline
		${\rm ^{146}Ba} $& $-$0.95 & $-$1.23 & $-$2.58 & $-$1.70 & 0.681 & 10.38 & 7.60 & 6.46\\
		${\rm ^{106}Mo}$ & $-$1.04 & $-$1.47 & $-$2.85 & $-$1.74 & 0.763 & 8.32 & 6.23 & 5.49\\ \hline
		${\rm ^{148}Ce}$ & $-$1.00 & $-$1.27 & $-$2.65 & $-$1.80 & 0.759 & 10.71 & 7.86 & 7.38\\
		${\rm ^{104}Zr}$ & $-$0.93 & $-$1.27 & $-$3.15 & $-$1.97 & 0.740 & 8.24 & 6.21 & 5.48\\ \hline
		${\rm ^{150}Ce}$ & $-$0.79 & $-$1.03 & $-$3.47 & $-$2.28 & 0.794 & 11.24 & 8.88 & 7.84\\
		${\rm ^{102}Zr}$ & $-$0.97 & $-$1.24 & $-$3.00 & $-$2.02 & 0.729 & 8.21 & 6.38 & 5.48 \\ \hline
		${\rm ^{152}Nd}$ & $-$0.77 & $-$1.05 & $-$3.77 & $-$2.39 & 0.857 & 11.77 & 9.47 & 8.33 \\
		${\rm ^{100}Sr}$ & $-$0.78 & $-$0.79 & $-$3.54 & $-$2.97 & 0.673 & 8.18 & 6.68 & 5.81 \\ \hline
		${\rm ^{154}Nd}$ & $-$0.69 & $-$0.62 & $-$4.21 & $-$4.03 & 0.864 & 11.70 & 10.72 & 8.37\\
		${\rm ^{98}Sr}$ & $-$0.73 & $-$1.30 & $-$3.66 & $-$1.76 & 0.645 & 7.69 & 6.53 & 5.05\\
		\end{tabular}
	\end{ruledtabular}
	\end{table*}
	
	\subsection{Optimal model search}\label{sbs:searching}
	
	With this information, it is finally possible to identify the most correct model for the moments of inertia. Since there are currently no experimental methods capable of estimating the moments of inertia of deformed FF, we will proceed indirectly. This method consists in comparing the mean values of the FF spins with their analogous experimental values taken from the work~\cite{wilson2021} for the case of spontaneous fission $\rm ^{252}Cf$. For this purpose we will use the apparatus proposed in the work~\cite{kadmensky2024}. Since the fragments emitted from the fissile nucleus in the vicinity of the scission point must be at ``cold" non-equilibrium states~\cite{kadmensky2005}, one should consider only zero oscillatory wave functions in the momentum representation~\cite{kadmensky2017, kadmensky2024} when constructing their mean values:
	\begin{equation}
		\begin{aligned}
			P &  \left(J_{k_x},J_{k_y} \right) = P (J_{k_x}) P(J_{k_y}) = \\
			& = \frac{1}{\pi I_k \hbar \omega _k} \exp \left[ - \frac{J_{k_x}^2+ J_{k_y}^2}{I_k \hbar \omega_k}\right],
		\end{aligned}
	\end{equation}
	where $k$ denotes the type of bending ($b$) or wriggling ($w$) oscillations, energies of the indicated zero oscillations~\cite{nix1965} are $\hbar {\omega }_w=2.5$ MeV; $\hbar {\omega }_w=0.9$ MeV, respectively. Using this form of the spin distributions and performing a number by simple transformations, one can derive expressions for calculating the mean value of the PFF spins:	
	\begin{equation} \label{eq:18}
		\overline{J_i}=\int\limits_{0}^{\infty }{{J_1}\frac{2{J_1}}{d_i}\exp \left[ \frac{J_{1}^{2}}{d_i} \right]d{J_1}}=\frac{\sqrt{\pi {d_i}}}{2},
	\end{equation}
	with coefficient $d_i$ is $\frac{I_i^2 I_w \hbar \omega_w + I_b \hbar \omega_b (I_1 + I_2)^2}{(I_1 + I_2)^2}$ and $i=\left( 1,2 \right)$ is fragment index.
	
\begin{figure}[h]
		\centering
		\includegraphics[width=\linewidth]{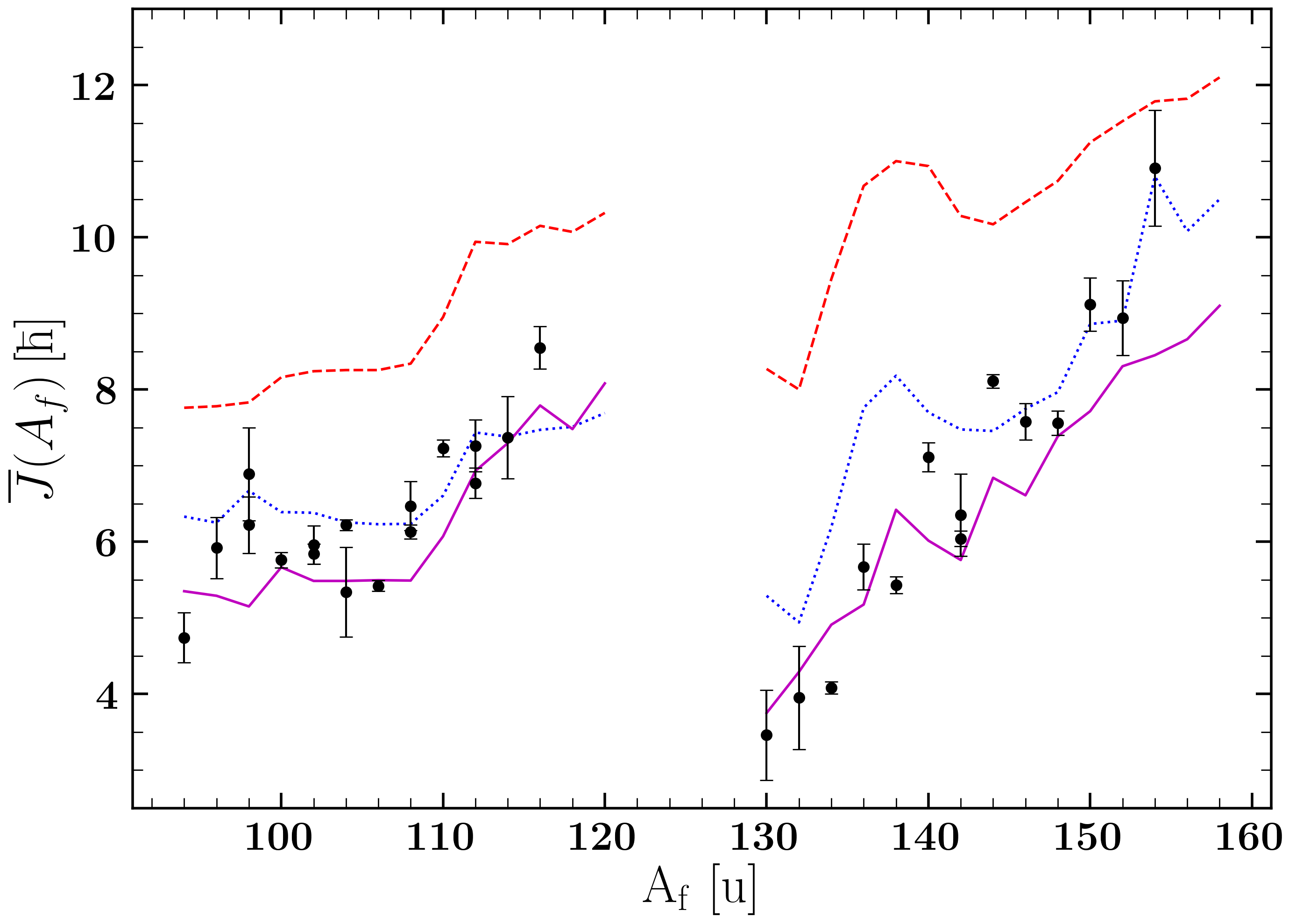}
		\vspace{-2em}
		\caption{The mean spin of FFs as a function of their mass for the spontaneous fission $\rm ^{252}Cf$, obtained using three estimates of the moments of inertia. In the superfluid approach, it is denoted by the red dashed line in the case of an oscillatory potential, and by the blue dashed line in the case of a rectangular potential, in the case of the hydrodynamic model - magenta line. Experimental points are taken from~\cite{wilson2021}.}\label{fig:5}
	\end{figure}

	Taking the formula~\eqref{eq:18}, then, we obtain estimates for three distinct models of moments of inertia, which are presented in~\cref{fig:5}.

	Comparing the obtained theoretical curves with the experimental data of the work~\cite{wilson2021}, a qualitative agreement can be observed only for the magenta line, i.e. when estimating the moments of inertia within the hydrodynamic model. In contrast, the mean values of the spins calculated in the superfluid approach: rectangular and oscillatory potentials (blue and red dashed lines) are in poorer agreement, especially for the latter.

	This apparently contradictory picture is extremely interesting, because it shows that as long as the FF have small quadrupole deformations slightly different from spherical symmetry, the Cooper pairing and the superfluid nucleon-nucleon correlations work well. Therefore, the best result will be obtained by using a superfluid model with an oscillatory potential as shown at~\cref{fig:2}. However, as soon as PFFs move into non-equilibrium deformations, where they reach abnormally large values (close to 1), the mean path length becomes smaller than the nucleon size, and thus the hydrodynamic model begins to prevail, in which the collective motion of the nucleons becomes potential, as can be seen from the analysis~\cref{fig:5,tab:2}.
	
	\section{Summary}\label{sec:summary}
	
	This study has provided a detailed examination of various theoretical models for describing the FF moments of inertia. By analyzing the data obtained, several key conclusions have emerged.
	
	The "direct" approach described in subsection~\ref{sbs:superfluid approach}, which applies a superfluid model of the nucleus with an oscillatory potential, demonstrates the closest agreement with experimental results. This model is particularly effective when considering small quadrupole deformations near spherical symmetry, where residual interactions, such as Cooper pairing and superfluid nucleon-nucleon correlations, play a significant role. Under these conditions, the average path length of nucleons is comparable to the size of the nucleus, further validating the model's applicability.
	
	In contrast, the "indirect" approach discussed in subsection~\ref{sbs:searching}, which compares models based on the average values of spins, proves less reliable. It only achieves agreement when employing a hydrodynamic model, which becomes dominant under conditions of substantial quadrupole deformation. Here, the path length of nucleons is smaller than the nucleus size, causing the hydrodynamic model to better representation of the fission dynamics.
	
	These findings highlight the necessity of applying different models depending on the nuclear deformation conditions. This distinction is critical not only for theoretical understanding but also for practical applications, such as in the development of computational tools like \texttt{FREYA}. The authors of FREYA, for example, used the "at hot" approximation for moments of inertia, which is less accurate compared to the approaches discussed in this study.
	
	The results presented in this article are unique, as they provide the first comparative analysis of FF moments of inertia using two distinct approaches. This work lays the groundwork for further research that should extend these findings across a broader range of nuclei, allowing for a more precise determination of each model's applicability. Such efforts will help to advance the boundaries of modern quantum fission theory.
	
	The study underscores the importance of understanding the nature of FF moments of inertia and their role in nuclear fission, contributing valuable insights to the field and opening avenues for future studies.
	
	\begin{acknowledgments}
	
	Authors are very grateful to Prof. S.G. Kadmensky for help and discussion in clarifying details on the topic of ideas about the ``coldness'' of the fissile nucleus, and to Prof. V.I. Furman for very useful comments and interesting discussion on the present work.
	
	\end{acknowledgments}
	
	\bibliographystyle{apsrev4-2}
	\bibliography{ref.bib}

\providecommand{\noopsort}[1]{}\providecommand{\singleletter}[1]{#1}%
\begin{thebibliography}{44}%
\makeatletter
\providecommand \@ifxundefined [1]{%
 \@ifx{#1\undefined}
}%
\providecommand \@ifnum [1]{%
 \ifnum #1\expandafter \@firstoftwo
 \else \expandafter \@secondoftwo
 \fi
}%
\providecommand \@ifx [1]{%
 \ifx #1\expandafter \@firstoftwo
 \else \expandafter \@secondoftwo
 \fi
}%
\providecommand \natexlab [1]{#1}%
\providecommand \enquote  [1]{``#1''}%
\providecommand \bibnamefont  [1]{#1}%
\providecommand \bibfnamefont [1]{#1}%
\providecommand \citenamefont [1]{#1}%
\providecommand \href@noop [0]{\@secondoftwo}%
\providecommand \href [0]{\begingroup \@sanitize@url \@href}%
\providecommand \@href[1]{\@@startlink{#1}\@@href}%
\providecommand \@@href[1]{\endgroup#1\@@endlink}%
\providecommand \@sanitize@url [0]{\catcode `\\12\catcode `\$12\catcode
  `\&12\catcode `\#12\catcode `\^12\catcode `\_12\catcode `\%12\relax}%
\providecommand \@@startlink[1]{}%
\providecommand \@@endlink[0]{}%
\providecommand \url  [0]{\begingroup\@sanitize@url \@url }%
\providecommand \@url [1]{\endgroup\@href {#1}{\urlprefix }}%
\providecommand \urlprefix  [0]{URL }%
\providecommand \Eprint [0]{\href }%
\providecommand \doibase [0]{https://doi.org/}%
\providecommand \selectlanguage [0]{\@gobble}%
\providecommand \bibinfo  [0]{\@secondoftwo}%
\providecommand \bibfield  [0]{\@secondoftwo}%
\providecommand \translation [1]{[#1]}%
\providecommand \BibitemOpen [0]{}%
\providecommand \bibitemStop [0]{}%
\providecommand \bibitemNoStop [0]{.\EOS\space}%
\providecommand \EOS [0]{\spacefactor3000\relax}%
\providecommand \BibitemShut  [1]{\csname bibitem#1\endcsname}%
\let\auto@bib@innerbib\@empty
\bibitem [{\citenamefont {Stetcu}\ \emph {et~al.}(2021)\citenamefont {Stetcu},
  \citenamefont {Lovell}, \citenamefont {Talou}, \citenamefont {Kawano},
  \citenamefont {Marin}, \citenamefont {Pozzi},\ and\ \citenamefont
  {Bulgac}}]{stetcu2021}%
  \BibitemOpen
  \bibfield  {author} {\bibinfo {author} {\bibfnamefont {I.}~\bibnamefont
  {Stetcu}}, \bibinfo {author} {\bibfnamefont {A.~E.}\ \bibnamefont {Lovell}},
  \bibinfo {author} {\bibfnamefont {P.}~\bibnamefont {Talou}}, \bibinfo
  {author} {\bibfnamefont {T.}~\bibnamefont {Kawano}}, \bibinfo {author}
  {\bibfnamefont {S.}~\bibnamefont {Marin}}, \bibinfo {author} {\bibfnamefont
  {S.~A.}\ \bibnamefont {Pozzi}},\ and\ \bibinfo {author} {\bibfnamefont
  {A.}~\bibnamefont {Bulgac}},\ }\href
  {https://doi.org/10.1103/PhysRevLett.127.222502} {\bibfield  {journal}
  {\bibinfo  {journal} {Phys. Rev. Lett.}\ }\textbf {\bibinfo {volume} {127}},\
  \bibinfo {pages} {222502} (\bibinfo {year} {2021})}\BibitemShut {NoStop}%
\bibitem [{\citenamefont {Randrup}\ and\ \citenamefont
  {Vogt}(2021)}]{randrup&vogt2021}%
  \BibitemOpen
  \bibfield  {author} {\bibinfo {author} {\bibfnamefont {J.}~\bibnamefont
  {Randrup}}\ and\ \bibinfo {author} {\bibfnamefont {R.}~\bibnamefont {Vogt}},\
  }\href {https://doi.org/10.1103/PhysRevLett.127.062502} {\bibfield  {journal}
  {\bibinfo  {journal} {Phys. Rev. Lett.}\ }\textbf {\bibinfo {volume} {127}},\
  \bibinfo {pages} {062502} (\bibinfo {year} {2021})}\BibitemShut {NoStop}%
\bibitem [{\citenamefont {Randrup}\ and\ \citenamefont
  {Vogt}(2009)}]{randrup2009}%
  \BibitemOpen
  \bibfield  {author} {\bibinfo {author} {\bibfnamefont {J.}~\bibnamefont
  {Randrup}}\ and\ \bibinfo {author} {\bibfnamefont {R.}~\bibnamefont {Vogt}},\
  }\href {https://doi.org/https://doi.org/10.1103/PhysRevC.80.024601}
  {\bibfield  {journal} {\bibinfo  {journal} {Phys. Rev. C}\ }\textbf {\bibinfo
  {volume} {80}},\ \bibinfo {pages} {024601} (\bibinfo {year}
  {2009})}\BibitemShut {NoStop}%
\bibitem [{\citenamefont {Verbeke}\ \emph {et~al.}(2018)\citenamefont
  {Verbeke}, \citenamefont {Randrup},\ and\ \citenamefont
  {Vogt}}]{verbeke2018}%
  \BibitemOpen
  \bibfield  {author} {\bibinfo {author} {\bibfnamefont {J.~M.}\ \bibnamefont
  {Verbeke}}, \bibinfo {author} {\bibfnamefont {J.}~\bibnamefont {Randrup}},\
  and\ \bibinfo {author} {\bibfnamefont {R.}~\bibnamefont {Vogt}},\ }\href
  {https://doi.org/10.1016/j.cpc.2017.09.006} {\bibfield  {journal} {\bibinfo
  {journal} {Comp. Phys. Comm.}\ }\textbf {\bibinfo {volume} {222}},\ \bibinfo
  {pages} {263} (\bibinfo {year} {2018})}\BibitemShut {NoStop}%
\bibitem [{\citenamefont {D{\o}ssing}\ and\ \citenamefont
  {Randrup}(1985{\natexlab{a}})}]{dossing1985_I}%
  \BibitemOpen
  \bibfield  {author} {\bibinfo {author} {\bibfnamefont {T.}~\bibnamefont
  {D{\o}ssing}}\ and\ \bibinfo {author} {\bibfnamefont {J.}~\bibnamefont
  {Randrup}},\ }\href {https://doi.org/10.1016/0375-9474(85)90178-2} {\bibfield
   {journal} {\bibinfo  {journal} {Nucl. Phys. A}\ }\textbf {\bibinfo {volume}
  {433}},\ \bibinfo {pages} {215} (\bibinfo {year}
  {1985}{\natexlab{a}})}\BibitemShut {NoStop}%
\bibitem [{\citenamefont {D{\o}ssing}\ and\ \citenamefont
  {Randrup}(1985{\natexlab{b}})}]{dossing1985_II}%
  \BibitemOpen
  \bibfield  {author} {\bibinfo {author} {\bibfnamefont {T.}~\bibnamefont
  {D{\o}ssing}}\ and\ \bibinfo {author} {\bibfnamefont {J.}~\bibnamefont
  {Randrup}},\ }\href {https://doi.org/10.1016/0375-9474(85)90179-4} {\bibfield
   {journal} {\bibinfo  {journal} {Nucl. Phys. A}\ }\textbf {\bibinfo {volume}
  {433}},\ \bibinfo {pages} {280} (\bibinfo {year}
  {1985}{\natexlab{b}})}\BibitemShut {NoStop}%
\bibitem [{\citenamefont {Vogt}\ and\ \citenamefont
  {Randrup}(2013)}]{vogt2013}%
  \BibitemOpen
  \bibfield  {author} {\bibinfo {author} {\bibfnamefont {R.}~\bibnamefont
  {Vogt}}\ and\ \bibinfo {author} {\bibfnamefont {J.}~\bibnamefont {Randrup}},\
  }\href {https://doi.org/10.1103/PhysRevC.87.044602} {\bibfield  {journal}
  {\bibinfo  {journal} {Phys. Rev. C}\ }\textbf {\bibinfo {volume} {87}},\
  \bibinfo {pages} {044602} (\bibinfo {year} {2013})}\BibitemShut {NoStop}%
\bibitem [{\citenamefont {Randrup}\ and\ \citenamefont
  {Vogt}(2014)}]{randrup2014}%
  \BibitemOpen
  \bibfield  {author} {\bibinfo {author} {\bibfnamefont {J.}~\bibnamefont
  {Randrup}}\ and\ \bibinfo {author} {\bibfnamefont {R.}~\bibnamefont {Vogt}},\
  }\href {https://doi.org/10.1103/PhysRevC.89.044601} {\bibfield  {journal}
  {\bibinfo  {journal} {Phys. Rev. C}\ }\textbf {\bibinfo {volume} {89}},\
  \bibinfo {pages} {044601} (\bibinfo {year} {2014})}\BibitemShut {NoStop}%
\bibitem [{\citenamefont {Bulgac}\ \emph {et~al.}(2021)\citenamefont {Bulgac},
  \citenamefont {Abdurrahman}, \citenamefont {Jin}, \citenamefont {Godbey},
  \citenamefont {Schunck},\ and\ \citenamefont {Stetcu}}]{bulgac2021}%
  \BibitemOpen
  \bibfield  {author} {\bibinfo {author} {\bibfnamefont {A.}~\bibnamefont
  {Bulgac}}, \bibinfo {author} {\bibfnamefont {I.}~\bibnamefont {Abdurrahman}},
  \bibinfo {author} {\bibfnamefont {S.}~\bibnamefont {Jin}}, \bibinfo {author}
  {\bibfnamefont {K.}~\bibnamefont {Godbey}}, \bibinfo {author} {\bibfnamefont
  {N.}~\bibnamefont {Schunck}},\ and\ \bibinfo {author} {\bibfnamefont
  {I.}~\bibnamefont {Stetcu}},\ }\href
  {https://doi.org/10.1103/PhysRevLett.126.142502} {\bibfield  {journal}
  {\bibinfo  {journal} {Phys. Rev. Lett.}\ }\textbf {\bibinfo {volume} {126}},\
  \bibinfo {pages} {142502} (\bibinfo {year} {2021})}\BibitemShut {NoStop}%
\bibitem [{\citenamefont {Bulgac}(2022)}]{bulgac2022}%
  \BibitemOpen
  \bibfield  {author} {\bibinfo {author} {\bibfnamefont {A.}~\bibnamefont
  {Bulgac}},\ }\href {https://doi.org/10.1103/PhysRevC.106.014624} {\bibfield
  {journal} {\bibinfo  {journal} {Phys. Rev. C}\ }\textbf {\bibinfo {volume}
  {106}},\ \bibinfo {pages} {014624} (\bibinfo {year} {2022})}\BibitemShut
  {NoStop}%
\bibitem [{\citenamefont {Vogt}\ and\ \citenamefont
  {Randrup}(2021)}]{vogt2021}%
  \BibitemOpen
  \bibfield  {author} {\bibinfo {author} {\bibfnamefont {R.}~\bibnamefont
  {Vogt}}\ and\ \bibinfo {author} {\bibfnamefont {J.}~\bibnamefont {Randrup}},\
  }\href {https://doi.org/10.1103/PhysRevC.103.014610} {\bibfield  {journal}
  {\bibinfo  {journal} {Phys. Rev. C}\ }\textbf {\bibinfo {volume} {103}},\
  \bibinfo {pages} {014610} (\bibinfo {year} {2021})}\BibitemShut {NoStop}%
\bibitem [{\citenamefont {Sitenko}\ and\ \citenamefont
  {Tartakovskii}(2014)}]{sitenko2014}%
  \BibitemOpen
  \bibfield  {author} {\bibinfo {author} {\bibfnamefont {A.}~\bibnamefont
  {Sitenko}}\ and\ \bibinfo {author} {\bibfnamefont {V.}~\bibnamefont
  {Tartakovskii}},\ }\href@noop {} {\emph {\bibinfo {title} {Lectures on the
  Theory of the Nucleus}}}\ (\bibinfo  {publisher} {Elsevier},\ \bibinfo {year}
  {2014})\BibitemShut {NoStop}%
\bibitem [{\citenamefont {Migdal}(1960)}]{migdal1960}%
  \BibitemOpen
  \bibfield  {author} {\bibinfo {author} {\bibfnamefont {A.}~\bibnamefont
  {Migdal}},\ }\href {http://jetp.ras.ru/cgi-bin/r/index/e/10/1/p176?a=list}
  {\bibfield  {journal} {\bibinfo  {journal} {Sov. Phys. JETP}\ }\textbf
  {\bibinfo {volume} {10}},\ \bibinfo {pages} {176} (\bibinfo {year}
  {1960})}\BibitemShut {NoStop}%
\bibitem [{\citenamefont {Bogoljubov}(1958)}]{bogoliubov1958}%
  \BibitemOpen
  \bibfield  {author} {\bibinfo {author} {\bibfnamefont {N.~N.}\ \bibnamefont
  {Bogoljubov}},\ }\href {http://jetp.ras.ru/cgi-bin/dn/e_007_01_0041.pdf}
  {\bibfield  {journal} {\bibinfo  {journal} {Sov. Phys. JETP}\ }\textbf
  {\bibinfo {volume} {7}},\ \bibinfo {pages} {41} (\bibinfo {year}
  {1958})}\BibitemShut {NoStop}%
\bibitem [{\citenamefont {Tolmachev}\ and\ \citenamefont
  {Tiablikov}(1958)}]{tolmachev1958}%
  \BibitemOpen
  \bibfield  {author} {\bibinfo {author} {\bibfnamefont {V.}~\bibnamefont
  {Tolmachev}}\ and\ \bibinfo {author} {\bibfnamefont {S.}~\bibnamefont
  {Tiablikov}},\ }\href {http://jetp.ras.ru/cgi-bin/dn/e_007_01_0046.pdf}
  {\bibfield  {journal} {\bibinfo  {journal} {Sov. Phys. JETP}\ }\textbf
  {\bibinfo {volume} {7}},\ \bibinfo {pages} {46} (\bibinfo {year}
  {1958})}\BibitemShut {NoStop}%
\bibitem [{\citenamefont {Lovchikova}\ \emph {et~al.}(1983)\citenamefont
  {Lovchikova}, \citenamefont {Maksyutenko}, \citenamefont {Simakov},\ and\
  \citenamefont {Trufanov}}]{lovchikova1983}%
  \BibitemOpen
  \bibfield  {author} {\bibinfo {author} {\bibfnamefont {G.}~\bibnamefont
  {Lovchikova}}, \bibinfo {author} {\bibfnamefont {B.}~\bibnamefont
  {Maksyutenko}}, \bibinfo {author} {\bibfnamefont {S.}~\bibnamefont
  {Simakov}},\ and\ \bibinfo {author} {\bibfnamefont {A.}~\bibnamefont
  {Trufanov}},\ }\href@noop {} {}\bibinfo {type} {Tech. Rep.}\ (\bibinfo
  {institution} {Gosudarstvennyj Komitet po Ispol'zovaniyu Atomnoj Energii
  SSSR},\ \bibinfo {year} {1983})\BibitemShut {NoStop}%
\bibitem [{\citenamefont {Abu El~Sheikh}\ \emph {et~al.}(2020)\citenamefont
  {Abu El~Sheikh}, \citenamefont {Okhunov1}, \citenamefont {Abu~Kassim},\ and\
  \citenamefont {Khandaker}}]{sheikh2020}%
  \BibitemOpen
  \bibfield  {author} {\bibinfo {author} {\bibfnamefont {M.~K.~M.}\
  \bibnamefont {Abu El~Sheikh}}, \bibinfo {author} {\bibfnamefont
  {A.}~\bibnamefont {Okhunov1}}, \bibinfo {author} {\bibfnamefont
  {H.}~\bibnamefont {Abu~Kassim}},\ and\ \bibinfo {author} {\bibfnamefont
  {M.}~\bibnamefont {Khandaker}},\ }\href@noop {} {\bibfield  {journal}
  {\bibinfo  {journal} {Chin. Phys. C}\ }\textbf {\bibinfo {volume} {44}},\
  \bibinfo {pages} {114107} (\bibinfo {year} {2020})}\BibitemShut {NoStop}%
\bibitem [{\citenamefont {Bohr}\ and\ \citenamefont
  {Wheeler}(1939)}]{bohr1939}%
  \BibitemOpen
  \bibfield  {author} {\bibinfo {author} {\bibfnamefont {N.}~\bibnamefont
  {Bohr}}\ and\ \bibinfo {author} {\bibfnamefont {J.~A.}\ \bibnamefont
  {Wheeler}},\ }\href {https://doi.org/10.1103/PhysRev.56.426} {\bibfield
  {journal} {\bibinfo  {journal} {Phys. Rev.}\ }\textbf {\bibinfo {volume}
  {56}},\ \bibinfo {pages} {426} (\bibinfo {year} {1939})}\BibitemShut
  {NoStop}%
\bibitem [{\citenamefont {Nix}\ and\ \citenamefont
  {Swiatecki}(1965)}]{nix1965}%
  \BibitemOpen
  \bibfield  {author} {\bibinfo {author} {\bibfnamefont {J.}~\bibnamefont
  {Nix}}\ and\ \bibinfo {author} {\bibfnamefont {W.}~\bibnamefont
  {Swiatecki}},\ }\href@noop {} {\bibfield  {journal} {\bibinfo  {journal}
  {Nucl. Phys. A}\ }\textbf {\bibinfo {volume} {71}},\ \bibinfo {pages} {1}
  (\bibinfo {year} {1965})}\BibitemShut {NoStop}%
\bibitem [{\citenamefont {Bohr}\ and\ \citenamefont
  {Mottelson}(1998)}]{bohr_mottelson}%
  \BibitemOpen
  \bibfield  {author} {\bibinfo {author} {\bibfnamefont {A.}~\bibnamefont
  {Bohr}}\ and\ \bibinfo {author} {\bibfnamefont {B.}~\bibnamefont
  {Mottelson}},\ }\href {https://books.google.co.uk/books?id=NNZQDQAAQBAJ}
  {\emph {\bibinfo {title} {Nuclear Structure (In 2 Volumes)}}}\ (\bibinfo
  {publisher} {World Scientific Publishing Company},\ \bibinfo {year}
  {1998})\BibitemShut {NoStop}%
\bibitem [{\citenamefont {Sushkov}\ and\ \citenamefont
  {Flambaum}(1982)}]{sushkov1982}%
  \BibitemOpen
  \bibfield  {author} {\bibinfo {author} {\bibfnamefont {O.~P.}\ \bibnamefont
  {Sushkov}}\ and\ \bibinfo {author} {\bibfnamefont {V.~V.}\ \bibnamefont
  {Flambaum}},\ }\href {https://doi.org/10.1070/PU1982v025n01ABEH004491}
  {\bibfield  {journal} {\bibinfo  {journal} {Sov. Phys.-Uspekhi}\ }\textbf
  {\bibinfo {volume} {25}},\ \bibinfo {pages} {1} (\bibinfo {year}
  {1982})}\BibitemShut {NoStop}%
\bibitem [{\citenamefont {Tanimura}\ and\ \citenamefont
  {Fliessbach}(1987)}]{tanimura1987}%
  \BibitemOpen
  \bibfield  {author} {\bibinfo {author} {\bibfnamefont {O.}~\bibnamefont
  {Tanimura}}\ and\ \bibinfo {author} {\bibfnamefont {T.}~\bibnamefont
  {Fliessbach}},\ }\href {https://doi.org/10.1007/BF01289634} {\bibfield
  {journal} {\bibinfo  {journal} {Z. Phys. A}\ }\textbf {\bibinfo {volume}
  {328}},\ \bibinfo {pages} {475} (\bibinfo {year} {1987})}\BibitemShut
  {NoStop}%
\bibitem [{\citenamefont {Barabanov}\ and\ \citenamefont
  {Furman}(1997)}]{barabanov1997}%
  \BibitemOpen
  \bibfield  {author} {\bibinfo {author} {\bibfnamefont {A.}~\bibnamefont
  {Barabanov}}\ and\ \bibinfo {author} {\bibfnamefont {W.}~\bibnamefont
  {Furman}},\ }\href {https://doi.org/10.1007/s002180050260} {\bibfield
  {journal} {\bibinfo  {journal} {Z. Phys. A}\ }\textbf {\bibinfo {volume}
  {357}},\ \bibinfo {pages} {411} (\bibinfo {year} {1997})}\BibitemShut
  {NoStop}%
\bibitem [{\citenamefont {Kadmensky}(2002)}]{kadmensky2002}%
  \BibitemOpen
  \bibfield  {author} {\bibinfo {author} {\bibfnamefont {S.}~\bibnamefont
  {Kadmensky}},\ }\href {https://doi.org/10.1134/1.1501650} {\bibfield
  {journal} {\bibinfo  {journal} {Phys. At. Nucl.}\ }\textbf {\bibinfo {volume}
  {65}},\ \bibinfo {pages} {1390} (\bibinfo {year} {2002})}\BibitemShut
  {NoStop}%
\bibitem [{\citenamefont {Kadmensky}(2004)}]{kadmensky2004}%
  \BibitemOpen
  \bibfield  {author} {\bibinfo {author} {\bibfnamefont {S.}~\bibnamefont
  {Kadmensky}},\ }\href {https://doi.org/10.1134/1.1648914} {\bibfield
  {journal} {\bibinfo  {journal} {Phys. At. Nucl.}\ }\textbf {\bibinfo {volume}
  {67}},\ \bibinfo {pages} {241} (\bibinfo {year} {2004})}\BibitemShut
  {NoStop}%
\bibitem [{\citenamefont {Kadmensky}(2005)}]{kadmensky2005}%
  \BibitemOpen
  \bibfield  {author} {\bibinfo {author} {\bibfnamefont {S.}~\bibnamefont
  {Kadmensky}},\ }\href {https://doi.org/10.1134/1.2149077} {\bibfield
  {journal} {\bibinfo  {journal} {Phys. At. Nucl.}\ }\textbf {\bibinfo {volume}
  {68}},\ \bibinfo {pages} {1968} (\bibinfo {year} {2005})}\BibitemShut
  {NoStop}%
\bibitem [{\citenamefont {Bunakov}\ \emph {et~al.}(2016)\citenamefont
  {Bunakov}, \citenamefont {Kadmensky},\ and\ \citenamefont
  {Lyubashevsky}}]{bunakov2016}%
  \BibitemOpen
  \bibfield  {author} {\bibinfo {author} {\bibfnamefont {V.}~\bibnamefont
  {Bunakov}}, \bibinfo {author} {\bibfnamefont {S.}~\bibnamefont {Kadmensky}},\
  and\ \bibinfo {author} {\bibfnamefont {D.}~\bibnamefont {Lyubashevsky}},\
  }\href {https://doi.org/10.1134/S1063778816020046} {\bibfield  {journal}
  {\bibinfo  {journal} {Phys. At. Nucl.}\ }\textbf {\bibinfo {volume} {79}},\
  \bibinfo {pages} {304} (\bibinfo {year} {2016})}\BibitemShut {NoStop}%
\bibitem [{\citenamefont {Kadmensky}\ \emph {et~al.}(2017)\citenamefont
  {Kadmensky}, \citenamefont {Bunakov},\ and\ \citenamefont
  {Lyubashevsky}}]{kadmensky2017}%
  \BibitemOpen
  \bibfield  {author} {\bibinfo {author} {\bibfnamefont {S.}~\bibnamefont
  {Kadmensky}}, \bibinfo {author} {\bibfnamefont {V.}~\bibnamefont {Bunakov}},\
  and\ \bibinfo {author} {\bibfnamefont {D.}~\bibnamefont {Lyubashevsky}},\
  }\href {https://doi.org/https://doi.org/10.1134/S106377881705012X} {\bibfield
   {journal} {\bibinfo  {journal} {Phys. At. Nucl.}\ }\textbf {\bibinfo
  {volume} {80}},\ \bibinfo {pages} {850} (\bibinfo {year} {2017})}\BibitemShut
  {NoStop}%
\bibitem [{\citenamefont {Carruthers}\ and\ \citenamefont
  {Nieto}(1968)}]{carruthers1968}%
  \BibitemOpen
  \bibfield  {author} {\bibinfo {author} {\bibfnamefont {P.}~\bibnamefont
  {Carruthers}}\ and\ \bibinfo {author} {\bibfnamefont {M.~M.}\ \bibnamefont
  {Nieto}},\ }\href {https://doi.org/10.1103/RevModPhys.40.411} {\bibfield
  {journal} {\bibinfo  {journal} {Rev. Mod. Phys.}\ }\textbf {\bibinfo {volume}
  {40}},\ \bibinfo {pages} {411} (\bibinfo {year} {1968})}\BibitemShut
  {NoStop}%
\bibitem [{\citenamefont {Jesinger}\ \emph {et~al.}(2002)\citenamefont
  {Jesinger}, \citenamefont {K{\"o}tzle}, \citenamefont {G{\"o}nnenwein},
  \citenamefont {Mutterer}, \citenamefont {von Kalben}, \citenamefont
  {Danilyan}, \citenamefont {Pavlov}, \citenamefont {Petrov}, \citenamefont
  {Gagarski}, \citenamefont {Trzaska} \emph {et~al.}}]{jesinger2002}%
  \BibitemOpen
  \bibfield  {author} {\bibinfo {author} {\bibfnamefont {P.}~\bibnamefont
  {Jesinger}}, \bibinfo {author} {\bibfnamefont {A.}~\bibnamefont
  {K{\"o}tzle}}, \bibinfo {author} {\bibfnamefont {F.}~\bibnamefont
  {G{\"o}nnenwein}}, \bibinfo {author} {\bibfnamefont {M.}~\bibnamefont
  {Mutterer}}, \bibinfo {author} {\bibfnamefont {J.}~\bibnamefont {von
  Kalben}}, \bibinfo {author} {\bibfnamefont {G.}~\bibnamefont {Danilyan}},
  \bibinfo {author} {\bibfnamefont {V.}~\bibnamefont {Pavlov}}, \bibinfo
  {author} {\bibfnamefont {G.}~\bibnamefont {Petrov}}, \bibinfo {author}
  {\bibfnamefont {A.}~\bibnamefont {Gagarski}}, \bibinfo {author}
  {\bibfnamefont {W.}~\bibnamefont {Trzaska}}, \emph {et~al.},\ }\href
  {https://doi.org/10.1134/1.1471264} {\bibfield  {journal} {\bibinfo
  {journal} {Phys. At. Nucl.}\ }\textbf {\bibinfo {volume} {65}},\ \bibinfo
  {pages} {630} (\bibinfo {year} {2002})}\BibitemShut {NoStop}%
\bibitem [{\citenamefont {Danilyan}\ \emph {et~al.}(2009)\citenamefont
  {Danilyan}, \citenamefont {Granz}, \citenamefont {Krakhotin}, \citenamefont
  {Mezei}, \citenamefont {Novitsky}, \citenamefont {Pavlov}, \citenamefont
  {Russina}, \citenamefont {Shatalov},\ and\ \citenamefont
  {Wilpert}}]{danilyan2009}%
  \BibitemOpen
  \bibfield  {author} {\bibinfo {author} {\bibfnamefont {G.}~\bibnamefont
  {Danilyan}}, \bibinfo {author} {\bibfnamefont {P.}~\bibnamefont {Granz}},
  \bibinfo {author} {\bibfnamefont {V.}~\bibnamefont {Krakhotin}}, \bibinfo
  {author} {\bibfnamefont {F.}~\bibnamefont {Mezei}}, \bibinfo {author}
  {\bibfnamefont {V.}~\bibnamefont {Novitsky}}, \bibinfo {author}
  {\bibfnamefont {V.}~\bibnamefont {Pavlov}}, \bibinfo {author} {\bibfnamefont
  {M.}~\bibnamefont {Russina}}, \bibinfo {author} {\bibfnamefont
  {P.}~\bibnamefont {Shatalov}},\ and\ \bibinfo {author} {\bibfnamefont
  {T.}~\bibnamefont {Wilpert}},\ }\href
  {https://doi.org/10.1016/j.physletb.2009.06.068} {\bibfield  {journal}
  {\bibinfo  {journal} {Phys. Lett. B}\ }\textbf {\bibinfo {volume} {679}},\
  \bibinfo {pages} {25} (\bibinfo {year} {2009})}\BibitemShut {NoStop}%
\bibitem [{\citenamefont {Danilyan}\ \emph {et~al.}(2010)\citenamefont
  {Danilyan}, \citenamefont {Klenke}, \citenamefont {Krakhotin}, \citenamefont
  {Novitsky}, \citenamefont {Pavlov},\ and\ \citenamefont
  {Shatalov}}]{danilyan2010}%
  \BibitemOpen
  \bibfield  {author} {\bibinfo {author} {\bibfnamefont {G.}~\bibnamefont
  {Danilyan}}, \bibinfo {author} {\bibfnamefont {J.}~\bibnamefont {Klenke}},
  \bibinfo {author} {\bibfnamefont {V.}~\bibnamefont {Krakhotin}}, \bibinfo
  {author} {\bibfnamefont {V.}~\bibnamefont {Novitsky}}, \bibinfo {author}
  {\bibfnamefont {V.}~\bibnamefont {Pavlov}},\ and\ \bibinfo {author}
  {\bibfnamefont {P.}~\bibnamefont {Shatalov}},\ }\href
  {https://doi.org/10.1134/S1063778810070045} {\bibfield  {journal} {\bibinfo
  {journal} {Phys. At. Nucl.}\ }\textbf {\bibinfo {volume} {73}},\ \bibinfo
  {pages} {1116} (\bibinfo {year} {2010})}\BibitemShut {NoStop}%
\bibitem [{\citenamefont {Valsky}\ \emph {et~al.}(2010)\citenamefont {Valsky},
  \citenamefont {Gagarski}, \citenamefont {Guseva}, \citenamefont {Krinitsin},
  \citenamefont {Petrov}, \citenamefont {Pleva}, \citenamefont {Sokolov},
  \citenamefont {Petrova}, \citenamefont {Zavarukhina},\ and\ \citenamefont
  {Kuzmina}}]{valsky2010}%
  \BibitemOpen
  \bibfield  {author} {\bibinfo {author} {\bibfnamefont {G.}~\bibnamefont
  {Valsky}}, \bibinfo {author} {\bibfnamefont {A.}~\bibnamefont {Gagarski}},
  \bibinfo {author} {\bibfnamefont {I.}~\bibnamefont {Guseva}}, \bibinfo
  {author} {\bibfnamefont {D.}~\bibnamefont {Krinitsin}}, \bibinfo {author}
  {\bibfnamefont {G.}~\bibnamefont {Petrov}}, \bibinfo {author} {\bibfnamefont
  {Y.~S.}\ \bibnamefont {Pleva}}, \bibinfo {author} {\bibfnamefont
  {V.}~\bibnamefont {Sokolov}}, \bibinfo {author} {\bibfnamefont
  {V.}~\bibnamefont {Petrova}}, \bibinfo {author} {\bibfnamefont
  {T.}~\bibnamefont {Zavarukhina}},\ and\ \bibinfo {author} {\bibfnamefont
  {T.}~\bibnamefont {Kuzmina}},\ }\href
  {https://doi.org/10.3103/S1062873810060080} {\bibfield  {journal} {\bibinfo
  {journal} {Bull. Rus. Acad. Sci: Phys}\ }\textbf {\bibinfo {volume} {74}},\
  \bibinfo {pages} {767} (\bibinfo {year} {2010})}\BibitemShut {NoStop}%
\bibitem [{\citenamefont {Gagarskii}\ \emph {et~al.}(2011)\citenamefont
  {Gagarskii}, \citenamefont {Guseva}, \citenamefont {Goennenwein},
  \citenamefont {Kopach}, \citenamefont {Mutterer}, \citenamefont {Kuz’mina},
  \citenamefont {Petrov}, \citenamefont {Tyurin},\ and\ \citenamefont
  {Nesvizhevsky}}]{gagarskii2011}%
  \BibitemOpen
  \bibfield  {author} {\bibinfo {author} {\bibfnamefont {A.}~\bibnamefont
  {Gagarskii}}, \bibinfo {author} {\bibfnamefont {I.}~\bibnamefont {Guseva}},
  \bibinfo {author} {\bibfnamefont {F.}~\bibnamefont {Goennenwein}}, \bibinfo
  {author} {\bibfnamefont {Y.~N.}\ \bibnamefont {Kopach}}, \bibinfo {author}
  {\bibfnamefont {M.}~\bibnamefont {Mutterer}}, \bibinfo {author}
  {\bibfnamefont {T.}~\bibnamefont {Kuz’mina}}, \bibinfo {author}
  {\bibfnamefont {G.}~\bibnamefont {Petrov}}, \bibinfo {author} {\bibfnamefont
  {G.}~\bibnamefont {Tyurin}},\ and\ \bibinfo {author} {\bibfnamefont
  {V.}~\bibnamefont {Nesvizhevsky}},\ }\href
  {https://doi.org/10.1134/S1063774511070133} {\bibfield  {journal} {\bibinfo
  {journal} {Cryst. Rep.}\ }\textbf {\bibinfo {volume} {56}},\ \bibinfo {pages}
  {1238} (\bibinfo {year} {2011})}\BibitemShut {NoStop}%
\bibitem [{\citenamefont {Guseva}\ \emph {et~al.}(2013)\citenamefont {Guseva},
  \citenamefont {Gagarski}, \citenamefont {Gusev}, \citenamefont {Petrov},\
  and\ \citenamefont {Valski}}]{guseva2013}%
  \BibitemOpen
  \bibfield  {author} {\bibinfo {author} {\bibfnamefont {I.}~\bibnamefont
  {Guseva}}, \bibinfo {author} {\bibfnamefont {A.}~\bibnamefont {Gagarski}},
  \bibinfo {author} {\bibfnamefont {Y.~I.}\ \bibnamefont {Gusev}}, \bibinfo
  {author} {\bibfnamefont {G.}~\bibnamefont {Petrov}},\ and\ \bibinfo {author}
  {\bibfnamefont {G.}~\bibnamefont {Valski}},\ }\href
  {https://doi.org/10.1134/S1547477113040067} {\bibfield  {journal} {\bibinfo
  {journal} {Phys. Part. Nucl. Let.}\ }\textbf {\bibinfo {volume} {10}},\
  \bibinfo {pages} {331} (\bibinfo {year} {2013})}\BibitemShut {NoStop}%
\bibitem [{\citenamefont {Danilyan}(2019)}]{danilyan2019}%
  \BibitemOpen
  \bibfield  {author} {\bibinfo {author} {\bibfnamefont {G.}~\bibnamefont
  {Danilyan}},\ }\href {https://doi.org/10.1134/S1063778819030050} {\bibfield
  {journal} {\bibinfo  {journal} {Phys. At. Nucl.}\ }\textbf {\bibinfo {volume}
  {82}},\ \bibinfo {pages} {250} (\bibinfo {year} {2019})}\BibitemShut
  {NoStop}%
\bibitem [{\citenamefont {Strutinsky}(1967)}]{strutinsky1967}%
  \BibitemOpen
  \bibfield  {author} {\bibinfo {author} {\bibfnamefont {V.}~\bibnamefont
  {Strutinsky}},\ }\href {https://doi.org/10.1016/0375-9474(67)90510-6}
  {\bibfield  {journal} {\bibinfo  {journal} {Nucl. Phys. A}\ }\textbf
  {\bibinfo {volume} {95}},\ \bibinfo {pages} {420} (\bibinfo {year}
  {1967})}\BibitemShut {NoStop}%
\bibitem [{\citenamefont {Hagmann}\ \emph {et~al.}(2016)\citenamefont
  {Hagmann}, \citenamefont {Verbeke}, \citenamefont {Vogt},\ and\ \citenamefont
  {Roundrup}}]{hagmann2016}%
  \BibitemOpen
  \bibfield  {author} {\bibinfo {author} {\bibfnamefont {C.}~\bibnamefont
  {Hagmann}}, \bibinfo {author} {\bibfnamefont {J.}~\bibnamefont {Verbeke}},
  \bibinfo {author} {\bibfnamefont {R.}~\bibnamefont {Vogt}},\ and\ \bibinfo
  {author} {\bibfnamefont {J.}~\bibnamefont {Roundrup}},\ }\href@noop {} {\emph
  {\bibinfo {title} {Fission Reaction Event Yield Algorithm}}},\ \bibinfo
  {type} {Tech. Rep.}\ (\bibinfo  {institution} {Lawrence Livermore National
  Laboratory (LLNL), Livermore, CA (United States)},\ \bibinfo {year}
  {2016})\BibitemShut {NoStop}%
\bibitem [{\citenamefont {Grudzevich}(2000)}]{grudzevich2000}%
  \BibitemOpen
  \bibfield  {author} {\bibinfo {author} {\bibfnamefont {O.~T.}\ \bibnamefont
  {Grudzevich}},\ }\href@noop {} {\bibfield  {journal} {\bibinfo  {journal}
  {Probl. At. Sci. Technol. Ser: Nucl. Const.}\ }\textbf {\bibinfo {volume}
  {1}},\ \bibinfo {pages} {39} (\bibinfo {year} {2000})}\BibitemShut {NoStop}%
\bibitem [{iae(1998)}]{iaea_handbook1998}%
  \BibitemOpen
  \href
  {https://www.iaea.org/publications/5331/handbook-for-calculations-of-nuclear-reaction-data-reference-input-parameter-library}
  {\emph {\bibinfo {title} {Handbook for Calculations of Nuclear Reaction Data
  Reference Input Parameter Library}}},\ \bibinfo {series} {TECDOC Series}\
  No.\ \bibinfo {number} {1034}\ (\bibinfo  {publisher} {IAEA},\ \bibinfo
  {address} {Vienna},\ \bibinfo {year} {1998})\BibitemShut {NoStop}%
\bibitem [{\citenamefont {Walsh}\ and\ \citenamefont
  {Boldeman}(1977)}]{walsh1977}%
  \BibitemOpen
  \bibfield  {author} {\bibinfo {author} {\bibfnamefont {R.}~\bibnamefont
  {Walsh}}\ and\ \bibinfo {author} {\bibfnamefont {J.}~\bibnamefont
  {Boldeman}},\ }\href {https://doi.org/10.1016/0375-9474(77)90378-5}
  {\bibfield  {journal} {\bibinfo  {journal} {Nucl. Phys. A}\ }\textbf
  {\bibinfo {volume} {276}},\ \bibinfo {pages} {189} (\bibinfo {year}
  {1977})}\BibitemShut {NoStop}%
\bibitem [{\citenamefont {M{\"o}ller}\ \emph {et~al.}(2016)\citenamefont
  {M{\"o}ller}, \citenamefont {Sierk}, \citenamefont {Ichikawa},\ and\
  \citenamefont {Sagawa}}]{moller2016}%
  \BibitemOpen
  \bibfield  {author} {\bibinfo {author} {\bibfnamefont {P.}~\bibnamefont
  {M{\"o}ller}}, \bibinfo {author} {\bibfnamefont {A.~J.}\ \bibnamefont
  {Sierk}}, \bibinfo {author} {\bibfnamefont {T.}~\bibnamefont {Ichikawa}},\
  and\ \bibinfo {author} {\bibfnamefont {H.}~\bibnamefont {Sagawa}},\ }\href
  {https://doi.org/10.1016/j.adt.2015.10.002} {\bibfield  {journal} {\bibinfo
  {journal} {At. Data Nucl. Data Tables}\ }\textbf {\bibinfo {volume} {109}},\
  \bibinfo {pages} {1} (\bibinfo {year} {2016})}\BibitemShut {NoStop}%
\bibitem [{\citenamefont {Wilson}\ \emph {et~al.}(2021)\citenamefont {Wilson},
  \citenamefont {Thisse}, \citenamefont {Lebois},\ and\ \citenamefont
  {et~al.}}]{wilson2021}%
  \BibitemOpen
  \bibfield  {author} {\bibinfo {author} {\bibfnamefont {J.}~\bibnamefont
  {Wilson}}, \bibinfo {author} {\bibfnamefont {D.}~\bibnamefont {Thisse}},
  \bibinfo {author} {\bibfnamefont {M.}~\bibnamefont {Lebois}},\ and\ \bibinfo
  {author} {\bibnamefont {et~al.}},\ }\href
  {https://doi.org/10.1038/s41586-021-03304-w} {\bibfield  {journal} {\bibinfo
  {journal} {Nature}\ }\textbf {\bibinfo {volume} {590}},\ \bibinfo {pages}
  {566} (\bibinfo {year} {2021})}\BibitemShut {NoStop}%
\bibitem [{\citenamefont {Kadmensky}\ \emph {et~al.}(2024)\citenamefont
  {Kadmensky}, \citenamefont {Lyubashevsky}, \citenamefont {Stepanov},\ and\
  \citenamefont {Pisklyukov}}]{kadmensky2024}%
  \BibitemOpen
  \bibfield  {author} {\bibinfo {author} {\bibfnamefont {S.}~\bibnamefont
  {Kadmensky}}, \bibinfo {author} {\bibfnamefont {D.}~\bibnamefont
  {Lyubashevsky}}, \bibinfo {author} {\bibfnamefont {D.}~\bibnamefont
  {Stepanov}},\ and\ \bibinfo {author} {\bibfnamefont {A.}~\bibnamefont
  {Pisklyukov}},\ }\href {https://doi.org/10.1134/S1063778824600155} {\bibfield
   {journal} {\bibinfo  {journal} {Phys. At. Nucl.}\ }\textbf {\bibinfo
  {volume} {87}},\ \bibinfo {pages} {359} (\bibinfo {year} {2024})}\BibitemShut
  {NoStop}%
\end{thebibliography}%

\end{document}